\documentclass[sigplan,10pt,nonacm]{acmart}
\pdfoutput=1
\setcopyright{none}
\acmConference{}{}{}

\startPage{1}

\usepackage{booktabs} 
                     
\usepackage{booktabs} 
\usepackage{tabularx}
\usepackage{graphicx}
\usepackage{amsmath}
\usepackage{amsthm}
\usepackage{subcaption}
\usepackage{multirow}
\usepackage{xspace}
\usepackage{listings}
\usepackage{times}
\usepackage{breqn}
\usepackage{algorithm}
\usepackage{algpseudocode}
\usepackage{amsmath}    
\usepackage{amssymb}    
\usepackage{amsfonts}
\usepackage{xcolor}
\usepackage{paralist}
\usepackage{tabulary}   
\usepackage[flushleft]{threeparttable}
\usepackage{bm}
\usepackage{csquotes}
\usepackage{rotating}
\usepackage{siunitx}
\usepackage{url}
\usepackage[T1]{fontenc}

\usepackage{textcomp}

\newcommand{\R}[1]{\textrm{#1}}

\newcommand{\B}[1]{\textbf{#1}}

\newcommand{\BU}[1]{\B{\underline{#1}}}

\newcommand{\SL}[1]{\S\ref{section:#1}}

\definecolor{swiftbuiltincolor}{rgb}{0,0,0}
\definecolor{swiftstringcolor}{rgb}{0,0,0}
\definecolor{swiftcommentcolor}{rgb}{0,0,0}

\newcommand{\PB}[1]{}

\newcommand{\eat}[1]{ }
\newcommand{\etal} {{et~al.}\xspace}

\newcommand{\Figure}[1]{Fig.~\ref{figure:#1}}
\newcommand{\TwoFigures}[2]{Figs.~\ref{figure:#1} and~\ref{figure:#2}}

\newcommand{\Table}[1]{Tbl.~\ref{table:#1}}

\newcommand{\Algorithm}[1]{Alg.~\ref{algorithm:#1}}
\newcommand{\Line}[1]{Line~\ref{line:#1}}
\newcommand{\TwoLines}[2]{Lines~\ref{line:#1}, \ref{line:#2}}
\newcommand{\MultiLines}[2]{Lines~\ref{line:#1}--\ref{line:#2}}

\usepackage[font={small,it},labelfont={small,it}]{subcaption}
\usepackage[labelfont={small,bf},font={small,bf}]{caption}
    \newcommand\ian[1]{}
    \newcommand\woz[1]{}
    \newcommand\yanfei[1]{}
    \newcommand\tong[1]{}

\begin{document}

\title[In-situ Workflow Auto-tuning]{\huge In-situ Workflow Auto-tuning via Combining Performance Models of Component Applications}

\author{Tong Shu}
\affiliation{
\institution{Southern Illinois University}
}
\email{tong.shu@siu.edu}

\author{Yanfei Guo}
\affiliation{
\institution{Argonne National Laboratory}
}
\email{yguo@anl.gov}

\author{Justin Wozniak}
\affiliation{
\institution{Argonne National Laboratory}
}
\email{woz@anl.gov}

\author{Xiaoning Ding}
\affiliation{
\institution{New Jersey Institute of Technology}
}
\email{xiaoning.ding@njit.edu}

\author{Ian Foster}
\affiliation{
\institution{Argonne Nat Lab and U.Chicago}
}
\email{foster@anl.gov}

\author{Tahsin Kurc}
\affiliation{
\institution{Stony Brook University}
}
\email{tahsin.kurc@stonybrook.edu}

\begin{abstract}
In-situ parallel workflows couple multiple component applications, 
such as simulation and analysis, via streaming data transfer 
in order to avoid data exchange via shared file systems.
Such workflows are challenging to configure for optimal performance
due to the large space of possible configurations.
Expert experience is rarely sufficient to identify optimal configurations,
and existing empirical auto-tuning approaches are inefficient
due to the high cost of obtaining training data for machine learning models.
It is also infeasible to optimize individual components independently,
due to component interactions.
We propose here a new auto-tuning method,
Component-based Ensemble Active Learning (CEAL),
that combines machine learning
techniques with knowledge of in-situ workflow structure to enable
automated workflow configuration with a limited number of performance measurements. 
Experiments with real applications demonstrate that CEAL can identify 
significantly better configurations than other approaches given compute time budgets.
For example, with 50 training samples, it reduces execution time and
computer time for a realistic workflow 
by 17.6\% and 40.8\% relative to random sampling, and by 12.4\% and 
32.5\% relative to a state-of-the-art algorithm, GEIST, respectively.
CEAL is also cost-effective: The tuned workflow need be run only 864 times
to pay off training sample collection costs,
40\% less than the 1444 times required with pure active learning.
\end{abstract}

\keywords{In-situ workflows, auto-tuning, active transfer learning, model combination}

\maketitle

\pagestyle{plain}

\vspace{-0.08in}
\section{Introduction}
\label{section:introduction}
\vspace{-0.05in}

Emerging scientific workflows couple simulation tasks with analysis, visualization, learning, and other data processing tasks.
It is increasingly infeasible to couple such workflows via file systems due to the performance gap between the computational and I/O components of HPC systems, as well as the
negative impacts on other users of the shared infrastructure.
In contrast, in-situ workflow solutions use network or shared memory to pass 
intermediate results~\cite{Ayachit:PerfAnalDesiConsAppExtscalSituInfra:SC16}.

Despite their advantages, in-situ workflows raise performance 
tuning challenges.
A single component application running in isolation can be tuned by  
selecting good configurations with known auto-tuning methods.
For example, empirical machine learning (ML) model-based auto-tuners have been 
widely applied to identify good configuration parameter values~\cite{Marathe:PerfMdlResConsTransLearn:SC17,
Yu:DatasizeAwareHighDimenConfAutoTunInMemClusterComp:ASPLOS18,
Ding:AutotunAlgoChoiInputSens:PLDI15}.
Multiple runs are performed with different parameter values, performance data 
from these runs are used to train the performance prediction model, 
and this model is used to identify a (close-to-)optimal configuration.

However, in the case of in-situ, multi-component workflows,
it is typically insufficient to tune each component independently because
components interact frequently and contend for resources during execution.
Parameter values that enable optimal performance for a component running in
isolation may lead to poor performance when the component is executed in the
workflow.
Ideally, all parameters from all components should be optimized together, but
that approach is rarely realistic with conventional methods, due to the multiplicative 
increase in potential parameter combinations.

A key issue for automated optimization of in-situ workflows is thus how to 
produce good results at an affordable cost. To this end,
we introduce here a new auto-tuning approach.
We leverage the workflow structure of the coupled
applications and combine the effectiveness of tuning a workflow as a
whole and the simplicity of
tuning individual components. We first train models on each component 
separately and then leverage the trained component models to guide 
the search for optimal parameter values for the workflow. 

We address two challenging issues. 
First, the component models of a workflow must be  %
combined into an integral model, which then can be used to help the search. 
We address this issue by leveraging high-level knowledge of the workflow structure
and how overall performance is affected 
by that of its components. 
We couple component models based on this knowledge to create a low-fidelity 
yet integral performance model of the workflow. 

Second, the integral model must be applied in the search for good
workflow parameter values, even though
the model cannot be used directly in the search due to its low fidelity.
Thus, we use it to help train a high-fidelity performance model of the workflow, 
which is then used in the search. 
Specifically, we use the low-fidelity model within an active learning (AL) 
algorithm to select good configurations as training samples. 
The intuition is that the search does not require a performance model with 
high fidelity in all circumstances. 
Instead, high fidelity is needed only when examining the parameter values close 
to the best configuration.

In auto-tuning, the dominant cost is that of running a workflow repeatedly with different 
configurations in order to collect enough data for accurate performance model generation. 
For conventional methods that treat an in-situ workflow as 
a whole and train its model in the same way as they do a single application,
the huge parameter space results in the workflow being run with many configurations.

Our solution substantially reduces costs in two ways. First,
each component application has a smaller parameter space than the whole
workflow; thus, we can create its low-fidelity model with 
relatively small training datasets. Second, the use of these low-fidelity
models in the active learning algorithm helps us to 
select whole-workflow training samples
that lead to a high-quality model with fewer whole-workflow runs.

This paper makes the following contributions: 
1) We propose and explore the idea of auto-tuning an in-situ worflow by 
combining structured performance models trained on component applications.
2) We implement the idea in a new in-situ workflow auto-tuning algorithm,
CEAL (\BU{C}omponent-based \BU{E}nsemble \BU{A}ctive \BU{L}earning).
3) We use three in-situ HPC workflows to experimentally verify the superiority 
of CEAL over other auto-tuning algorithms.
With just 25 training samples, 
CEAL can reduce computer time by 12--48\%.

\vspace{-0.08in}
\section{Background and Motivation}
\label{section:background}
\vspace{-0.05in}

HPC applications are often run repeatedly on similar computers and problems.
These similarities can make it rewarding to tune configuration parameters to improve performance.
Given the growing complexity of applications and HPC infrastructures, 
tuning increasingly
relies on auto-tuners, particularly empirical model-based auto-tuners, which train a 
performance model of an
application that they then use to search for 
and select a good set of configuration parameters.

Though there are various auto-tuning approaches~\cite{Balaprakash:AutotunHPCApp:PIEEE18},
this paper focuses on empirical model-based auto-tuning, because of its effectiveness and 
prevalence. For brevity, we here refer to
``empirical 
model-based auto-tuning/auto-tuners'' as ``auto-tuning/auto-tuners.'' 
 We next describe the basic mechanisms used in auto-tuners and
identify the challenges in designing such auto-tuners for in-situ workflows.

\vspace{-0.06in}
\subsection{Empirical Model-based Auto-tuners}
\vspace{-0.03in}

An auto-tuner typically has three components: collector, modeler, and searcher
\cite{Marathe:PerfMdlResConsTransLearn:SC17,
Thiagarajan:BootParaSpacExplFastTun:ICS18,
Golovin:GoogleVizierServBlacBoxOpt:KDD17,
Tillet:InputAwareAutoTunCompBoundHPCKernel:SC17,
Cao:UnderBlackAutoTunCompAnalStorSys:ATC18,
Meng:PatternAlgoAutotunGraphProcGPU:PPoPP19,
Behzad:OptIOPerfHPCAppAutotun:TOPC19}.
The \emph{collector} runs the target application with different configurations 
selected by the modeler, and collects performance measurements. 
The \emph{modeler} selects configurations from the parameter configuration 
space of the target application, drives the collector to obtain the 
corresponding performance measurements, and uses the measurements as training 
data to construct a surrogate performance model:
a high-dimensional function of configuration parameters, 
usually obtained by ML.
The \emph{searcher} uses the model to search for a good configuration, i.e., 
one that produces good performance.
During the search, the searcher uses the model to predict the performance
for the configurations being examined, and selects the configuration with
the best predicted performance.
The core of an auto-tuner design is an auto-tuning algorithm. 
Factors to consider when designing this algorithm include model 
type and the methods used to select configurations used as training samples, train
the model, and to search configuration space. 
Thus, for example, neural networks, which require many 
training samples, are not used in our solution, 
because the cost of collecting so many samples is prohibitive for HPC
workflows.

A well-designed auto-tuning algorithm can substantially both reduce the cost and 
improve the performance of an auto-tuner. For resource-intensive applications, 
such as HPC programs, {\it auto-tuner cost} is dominated by the time required to run 
the target application repeatedly to collect training data. Model training and configuration space search, in contrast, are inexpensive: for traditional ML models, 
such as boosted trees and random forests, they may take only 
a few minutes.  {\it Auto-tuner performance}, the capacity to find
good configurations and improve application performance,
is determined primarily by how well the model predicts 
application performance for given configurations. (It is also affected by 
how the parameter space is searched, but as search mechanisms are 
mature, most efforts focus on improving the model.)

\vspace{-0.06in}
\subsection{Auto-Tuning for In-situ Workflows}
\label{section:bgworkflows}
\vspace{-0.03in}

\begin{figure}[!t]
\begin{center}
  \begin{subfigure}[t]{.57\linewidth}
    \includegraphics[width=1.0\linewidth]{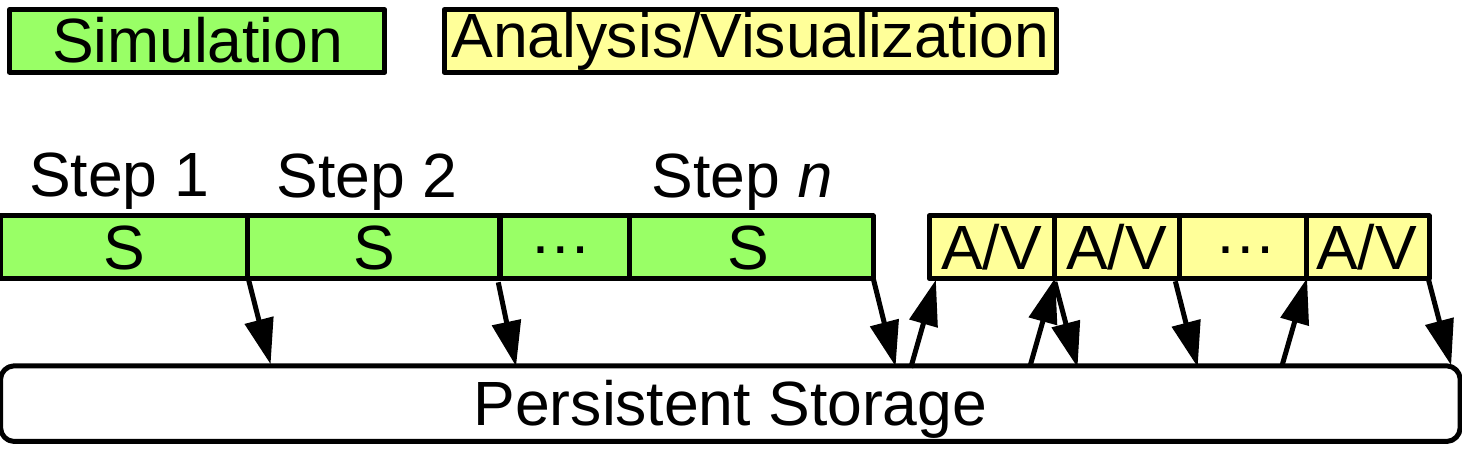}
    \caption{\it Post-hoc processing}
    \label{figure:post-hoc}
  \end{subfigure}
  \begin{subfigure}[t]{.41\linewidth}
    \includegraphics[width=1.0\linewidth]{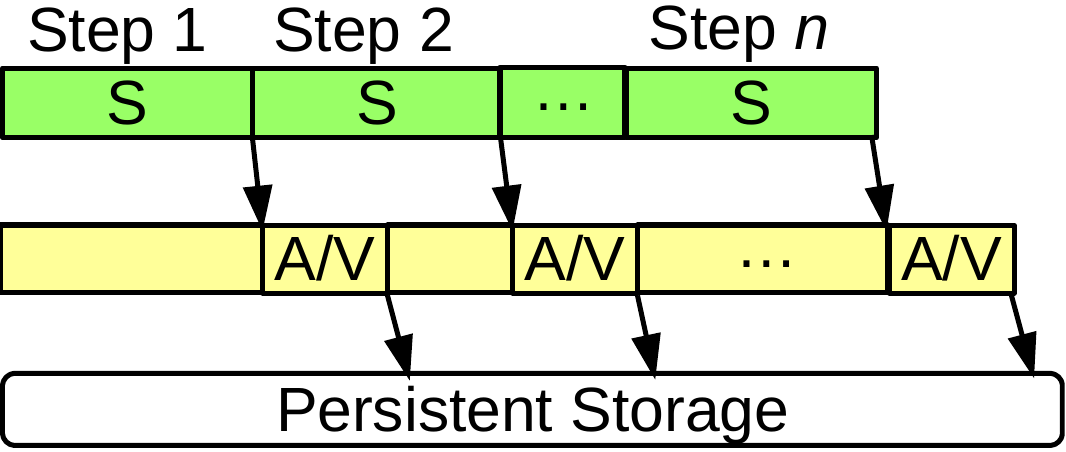}
    \caption{\it In-situ workflow}
    \label{figure:in-transit}
  \end{subfigure}
  \vspace{-3ex}
  \caption{Post hoc processing vs.\ in-situ workflows}
  \label{figure:processing}
\end{center}
\vspace{-0.25in}
\end{figure}

In our context, a workflow is %
a directed acyclic graph (DAG), with application components as nodes and data
as edges.
Components are typically coupled by using a higher-level programming model or language,
thus exposing a \mbox{\emph{structure}} that can be exploited for %
performance modeling and optimization.
In a file-based post-hoc processing workflow, component applications are executed in an order
determined by their data dependencies.
For example, in \Figure{post-hoc}, when the simulation 
finishes, it saves data to persistent storage;
only then can
the analysis/visualization, which processes 
the data, be started. Performance optimization can then proceed in two steps.
First, we model 
the performance 
of each component independently. Second, we find the best configuration
based on the DAG and the component models. %

In contrast, as shown in \Figure{in-transit}, component applications in an in-situ workflow 
run concurrently, exchanging data via network or 
shared memory. 
Workflow performance is determined by the complex interplay of the 
applications, which may involve factors such as load imbalances, 
contending network bandwidth, synchronizations, and locks~\cite{Fu:PerfAnalOptSituIntSimuAnalZipAppUp:HPDC18}.
High performance requires that the component applications execute in 
a balanced and coordinated way. 

For an in-situ workflow, performance optimization cannot simply be done
with separate performance models of each component application
due to the complicated interactions among component applications.
Instead, a performance model for the whole workflow must be built, and all 
parameters from all components should be optimized together. 
For empirical model-based auto-tuners, the fact that an in-situ workflow 
includes multiple coupled components raises considerable challenges. 
Because all parameters from all components must be considered together, 
the potential parameter combinations increases multiplicatively.
For example, in the two-component workflows of \SL{benchmarks}, 
the configuration space sizes are more than 10$^5\times$ larger than those of their
component applications.
This dramatically raises the number of configurations to be measured as
training samples for building a usable surrogate model. 
However, for in-situ workflows, it is not realistic to measure many
parameter combinations, given the high resource consumption of running 
HPC applications. 

Without fundamentally renovating auto-tuning algorithms, particularly the
techniques used to build the surrogate model, auto-tuners for in-situ 
workflows face a difficult dilemma---whether to suffer a prohibitive cost in 
creating the accurate surrogate model needed for optimal performance, 
or to tolerate the poor performance associated with an inaccurate surrogate 
model generated at an affordable cost.

The CEAL algorithm proposed in this paper fundamentally
improves the techniques to build surrogate models, such that auto-tuner cost  
can be reduced substantially while retaining high performance. 
The large resources needed to run a complete workflow repeatedly
when building an auto-tuner, data collection costs are usually limited in practical settings by a resource budget.
Thus, in this scenario, the advantage of the algorithm is reflected by 
improving the performance of the auto-tuner within a cost budget.

\vspace{-0.05in}
\section{Overview of the CEAL Approach}
\label{section:overview}
\vspace{-0.05in}

We now introduce the
methods that CEAL uses to identify well-performing configurations for in-situ workflows,
by building surrogate models with limited sampling costs.

\vspace{-0.05in}
\subsection{Basic Idea}
\vspace{-0.05in}

As auto-tuning cost is dominated by the collection of training samples,
we must select training samples carefully
and use them effectively, instead of selecting training
samples indiscriminately and extensively (e.g., by random sampling). 
The general idea of intelligent sampling has been explored in
different ways in ML and in auto-tuner designs~\cite{Mametjanov:AutotunFPGAParaPerfPow:FCCM15,
Behzad:OptIOPerfHPCAppAutotun:TOPC19, Thiagarajan:BootParaSpacExplFastTun:ICS18,
Marathe:PerfMdlResConsTransLearn:SC17}.
Our work is distinguished by how we exploit the workflow structure (\S\ref{section:bgworkflows}) to develop
special techniques that are particularly effective for auto-tuning 
in-situ workflows. 

Our approach leverages the following two characteristics of in-situ workflows: 1) An 
in-situ workflow consists of multiple components, which can run 
independently and may be reused across workflows. 2) The 
interaction among components means that if any component %
performs poorly, the workflow is unlikely to achieve high performance.

These characteristics suggest that we should avoid collecting samples in which 
a workflow performs poorly, as such samples are unlikely to help with finding 
well-performing configurations. 
But how are we to avoid collecting poor-performing samples in the absence of a performance model, which is why we want those samples in the first place?
We employ two ideas.
1) Leveraging the first characteristic, we build performance models for 
individual component applications.
Because the parameter spaces of component applications are much smaller than that
of the in-situ workflow, these models can be built at low cost, i.e., with only a few component application
runs. (Component reuse
across workflows can allow for the reuse of their models, further 
lowering costs.) 
2) Leveraging the second characteristic, plus the component models that we have just developed, we build a simple low-fidelity model that we use to guide our search
for well-performing configurations for the whole in-situ 
workflow, and focus sample collection on these configurations.

We also investigated other techniques that improve the selection of training samples
and thus may be integrated in our solution. Active learning iteratively uses the model
that is being refined to identify configurations that may lead to good performance, and 
focuses sample collection on
those configurations~\cite{Mametjanov:AutotunFPGAParaPerfPow:FCCM15,
Behzad:OptIOPerfHPCAppAutotun:TOPC19}.
Our approach similarly focuses on well-performing configurations,
as we discuss in \SL{actLearn}. 
We also explored the use of configurations with much smaller inputs (i.e., much
lower cost) for model building~\cite{Marathe:PerfMdlResConsTransLearn:SC17}. However, we found that programs often display quite different behaviors if input sizes (and problem 
sizes) change considerably, as when a reduced working set, corresponding to a
smaller input, can fit into memory or cache. Thus, 
this technique cannot be used in general when auto-tuning in-situ workflows.

\begin{figure}[!t]
\begin{center}
\includegraphics[width=.9\linewidth]{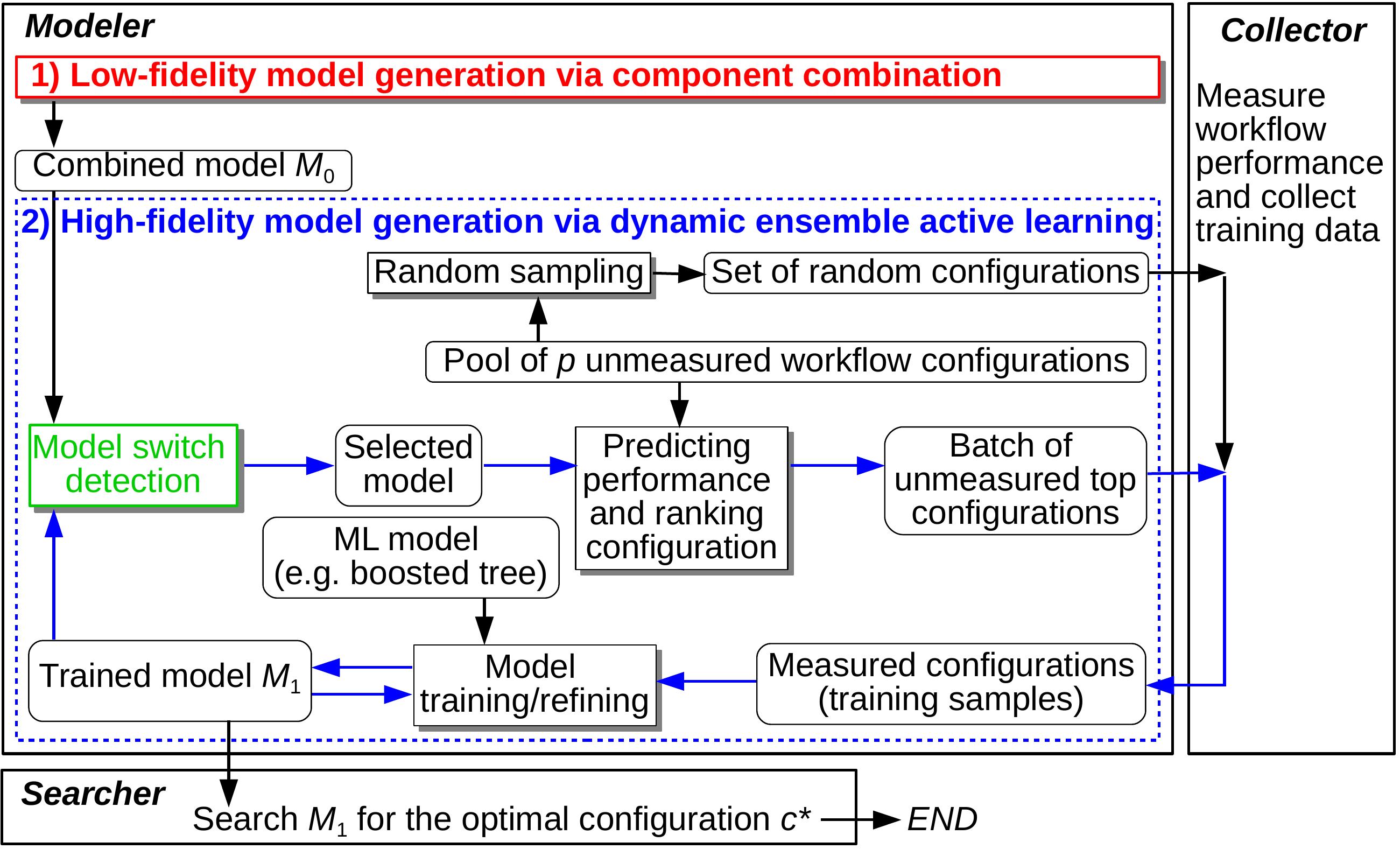}
\vspace{-2ex}
\caption{Structure of CEAL auto-tuning.}
\vspace{-0.4in}
\label{figure:CEAL}
\end{center}
\end{figure}

\vspace{-0.05in}
\subsection{The CEAL Algorithm Introduced}
\label{section:algoDesign}
\vspace{-0.05in}

To implement these ideas, we design the 
\BU{C}omponent-based \BU{E}nsemble \BU{A}ctive \BU{L}earning (CEAL)
auto-tuning algorithm 
shown in~\Figure{CEAL}.
CEAL consists of two main phases.

In the \B{Low-fidelity Model Generation via Component Combination} phase,
the red rectangle in~\Figure{CEAL}, we 
build performance models for a workflow's component applications
which we combine to form a simple yet integral workflow model. 
This simplicity means that the integral model can be obtained at 
low cost but yields only approximate predictions. %
We use this \emph{low-fidelity} workflow model ($M_0$ in \Figure{CEAL}) in the second phase, to evaluate 
configurations.

In the \B{High-fidelity Model Generation via Dynamic Ensemble Active Learning} phase, the orange dashed rectangle in~\Figure{CEAL},
we use a series of samples selected based on low-fidelity model scoring
to establish and improve a second, high-fidelity model
of the workflow ($M_1$ in \Figure{CEAL}). 
This is the surrogate model 
that the searcher will use to predict workflow performance so as to 
find an ideal configuration.

The high-fidelity model is primitive when first established, but keeps
evolving as more samples are collected and used in training, and 
may become a better choice for evaluating configurations than the 
low-fidelity model. 
Thus, we use a {\it model switch detection module} to monitor the two
models, and switch to using the high-fidelity
model to evaluate configurations when it becomes a better choice.
We stop evolving the high-fidelity
model when the cost budget is reached (i.e., we have tested a preset number of configurations). 

We introduce these two phases in the next two sections,
after which we present the detailed algorithm.

\vspace{-0.05in}
\section{Generating the Low-fidelity Model}
\label{section:combo}
\vspace{-0.05in}

We use the low-fidelity model to evaluate configurations by predicting
how well the workflow may perform. The main challenge is to build a usable 
model with minimal cost, such that the auto-tuner can quickly start from it and use
it to build and improve the high-fidelity model. 

\begin{figure}[!t]
\begin{center}
\includegraphics[width=.9\linewidth]{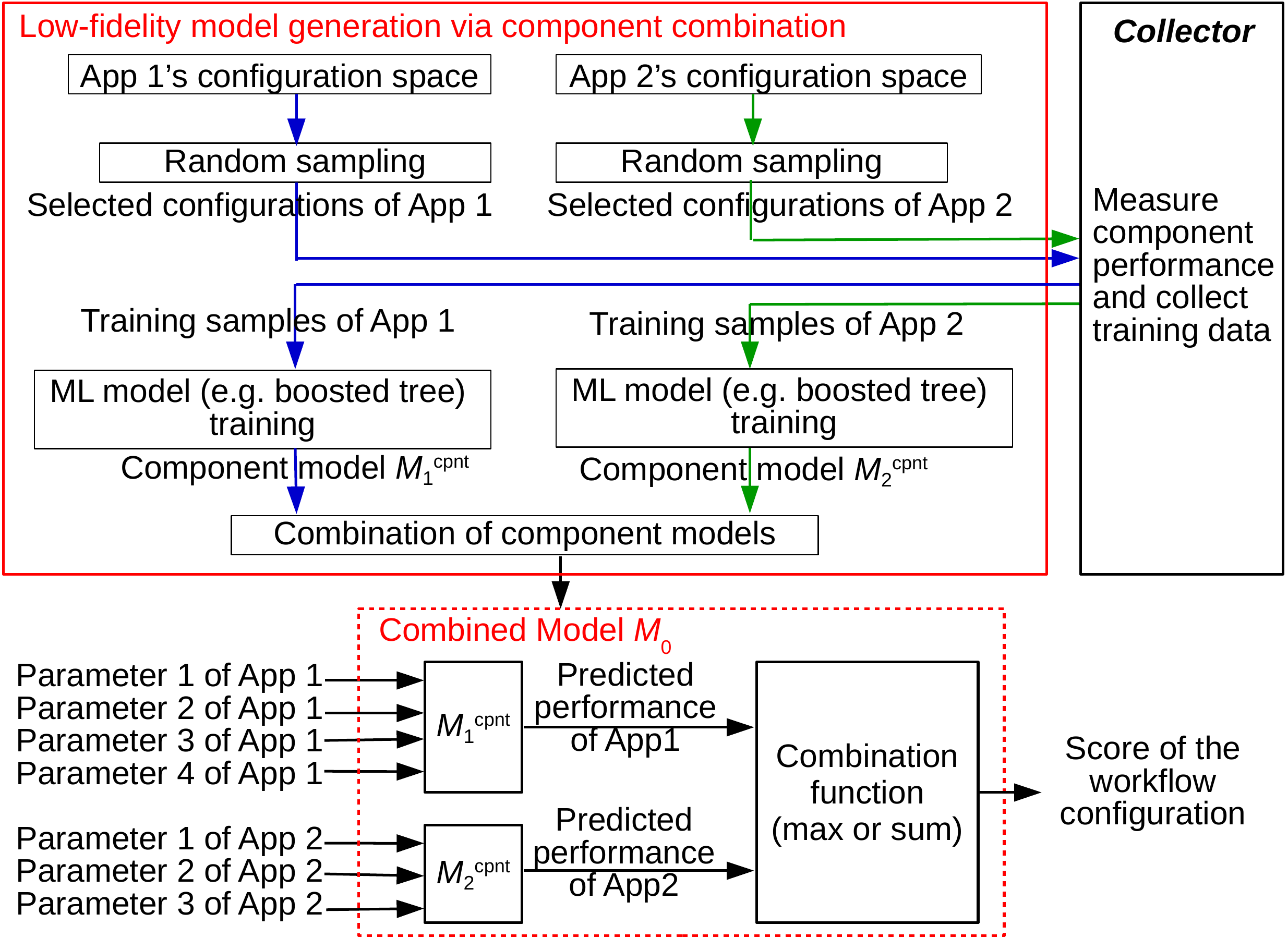}
\vspace{-2ex}
\caption{Combining component models.}
\vspace{-0.2in}
\label{figure:Combo}
\end{center}
\end{figure}

We build the low-fidelity model by first constructing and then combining predictive models 
for the component applications: see~\Figure{Combo}. 
Each individual \emph{component model}, $M_j$ ($j = 1, 2, \cdots$), outputs a
performance prediction for its component for a given configuration. Its predictions
should be aligned with the optimization goal of the auto-tuner; for example, 
execution times if the goal is shortening execution times. Component models 
can be built by using conventional methods, e.g., by randomly selecting configurations to 
collect samples and then training a boosted tree ML model. 
Since component applications are independent, they may be used separately
or in other workflows. Costs can be reduced by including measurements collected in earlier 
runs in training, or by reusing models developed for other workflows.
Due to space limitations, we do not elaborate here on how 
to build or reuse these models, but focus on 
how to combine component models to form the integral 
low-fidelity model.

There are two issues in forming the low-fidelity model. The first is what the 
model should output. As we will use this model only to choose among configurations,
we do not need it to predict workflow performance directly.
Instead, we make it output for each configuration just a score 
indicating how well the workflow performs relative to other configurations.

The second issue is how to combine per-component model results to build a low-fidelity model with minimal cost.  
One approach, which for later quantitative comparison we implement in an algorithm called ALpH, 
is to train a component-combining model $M_0$ from both component model predictions
and actual workflow runs. %
That is, for each candidate configuration $c$, we use the $\{ M_j \}$ to predict the performances, $\{ P_j \}$, of the
components for $c$; run the workflow with $c$, measuring its performance $p$; and add $\{ c, \{ P_j \}, p \}$ to the training set for $M_0$.
ALpH uses AL~\cite{Mametjanov:AutotunFPGAParaPerfPow:FCCM15} to select the configurations for which it generates such workflow training samples.

A deficiency of this approach is that it does not exploit any knowledge of the workflow structure.
An alternative, which we use in CEAL, is to use a simple function (e.g., max, min, sum),
chosen according to the performance metric being optimized by the auto-tuner,
to combine the component model predictions.
We select this function as follows.
If the performance metric is determined largely by the bottleneck components, 
such as execution time and throughput, we use {\it max} (for execution time)
or {\it min} (for throughput). If the performance metric
is largely an aggregation of the shares from all components, such as computing 
resource and energy consumption, we use {\it sum}. 
Notice that CEAL, unlike ALpH, does not need to run the workflow. 
We examine the relative accuracy and costs of CEAL and ALpH below.

We postpone detailed evaluation to \S\ref{section:evaluation}, but report here on a study in which we
use two optimization objectives with different metrics---shortening execution time 
and minimizing computer time---to illustrate and characterize the function approach. 
We define execution time as
wall-clock time and computer time as the number of core-hours 
consumed by workflow execution. 
We define the functions used to determine a configuration's scores as follows.
\vspace{-1ex}
\begin{equation}
\emph{Score}^{\R{e}}(c) = \max\limits_{j}{t^{\R{e}}}(c_j), 
\label{equation:max}
\vspace{-1ex}
\end{equation}

\noindent
\vspace{-1ex}
\begin{equation}
\emph{Score}^{\R{c}}(c) = \sum\limits_{j}{t^{\R{c}}}(c_j), 
\label{equation:sum}
\vspace{-1ex}
\end{equation}
where for a configuration $c$, $\emph{Score}^{\R{e}}(c)$ and $\emph{Score}^{\R{c}}(c)$ are 
the {\it \B{e}}xecution and {\it \B{c}}omputer times 
(the lower, the better) of configuration $c$; 
$c_j$ is the parameter values related to component $j$ 
extracted from $c$; and
$t^{\R{e}}(c_j)$ and $t^{\R{c}}(c_j)$ are the model-predicted execution and computer times of 
the $j^{\R{th}}$ component.

\begin{figure}[t]
\begin{center}
\includegraphics[width=.6\linewidth]{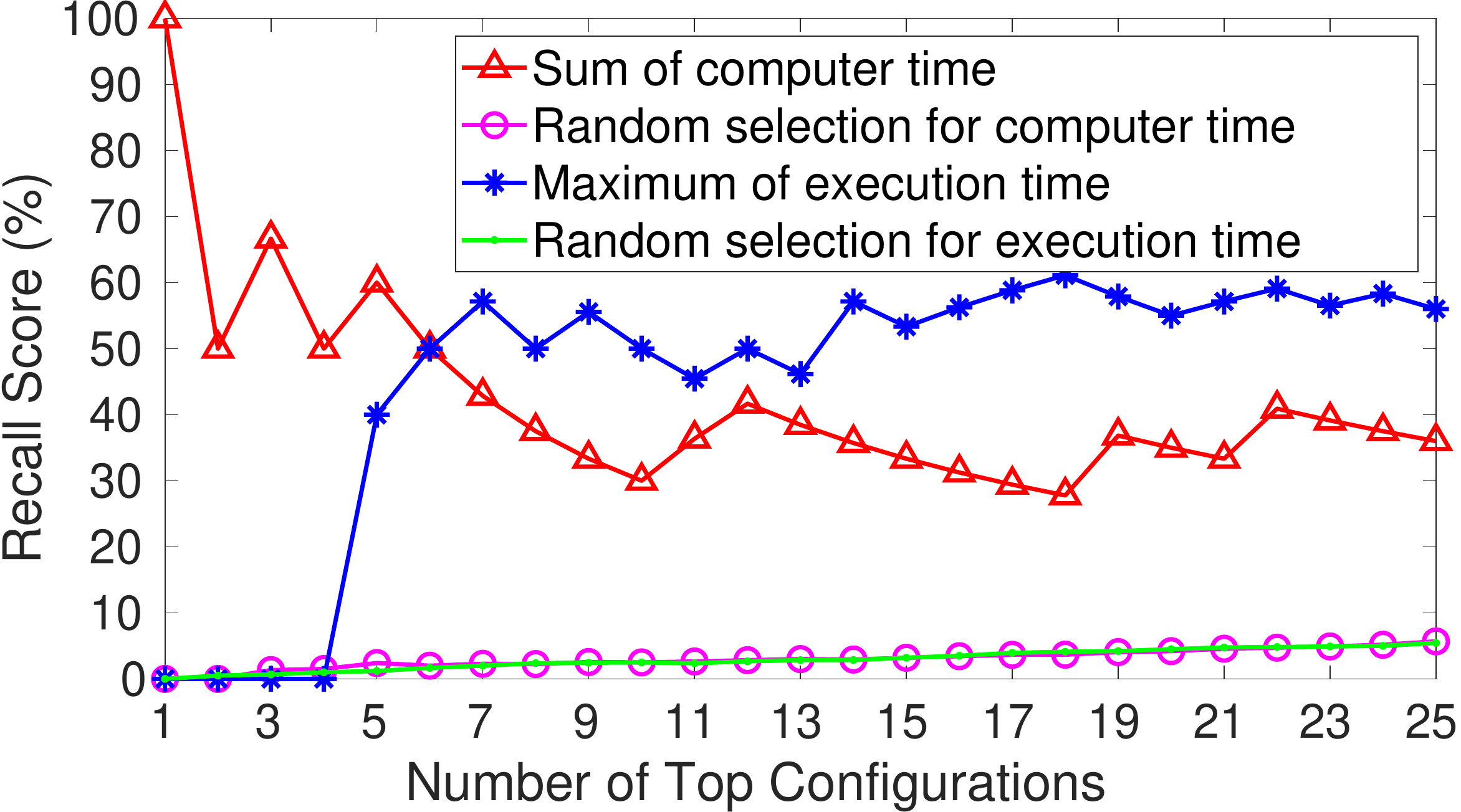}
\vspace{-2ex}
\caption{Recall scores based on combination functions.}\label{figure:cmb-rs}
\end{center}

\vspace{-5ex}

\end{figure}

To illustrate the effectiveness of this approach, \Figure{cmb-rs} shows the recall scores of 
the low-fidelity models (Eqns~\ref{equation:max} and~\ref{equation:sum}) 
when used to score 500 randomly selected configurations for workflow {\it LV}. 
(For details on 
experimental settings, see \S\ref{section:evaluation}.)
Recall scores reflect the possibility that the highest-scoring configurations lead to high workflow 
performance (\SL{recallScore}). To calculate recall scores, we rank configurations 
based both on their model-predicted scores and on
the performance observed when the workflow is executed with them.
The recall score of the top $n$ configurations is then the ratio between 1) the number of 
common configurations found in the top $n$ configurations on these ranked lists and 2) $n$.
We see that the models achieve recall scores above 30\% for top 5 to 25 
configurations, much higher than those of random selection. 
This verifies that even using simple functions in the low-fidelity model can 
effectively locate good configurations.

The solution described in this section mainly targets 
loosely-coupled in-situ workflows~\cite{Fu:PerfAnalOptSituIntSimuAnalZipAppUp:HPDC18,
Subedi:StackerAutoDataMoveEngiExtScalStagSituWf:SC18}, 
in which components are coupled via high-level libraries (e.g.,
ADIOS~\cite{Liu:ADIOSChallLessDevLeadClasIOFw:CCPE14}, 
Flexpath~\cite{Dayal:FlexpathTypePubSubSysLScalScieAnal:CCGrid14}, 
DataSpaces~\cite{Docan:DataSpacesIntCoorFwCoupWf:CC12}, 
FlexIO~\cite{Zheng:FlexIOMiddLocaFlexScieDataAnal:IPDPS13}, 
GLEAN~\cite{Vishwanath:TopoDataMoveStagIOAcceBlueGeneP:SC11}, 
DIMES \cite{Zhang:DIMESInMemStagDataTaskPlacCoupWf:CCPE17}, 
Decaf~\cite{Dreher:DecafDecoDataSituWf:TR17}, 
or Zipper~\cite{Fu:PerfAnalOptSituIntSimuAnalZipAppUp:HPDC18}).
The relative ease of development and deployment of loosely-coupled
in-situ workflows relative to tightly-coupled in-situ workflows makes the former much more prevalent. However, our 
solution can  easily be adapted to optimize tightly-coupled in-situ workflows.

\vspace{-0.05in}
\section{Generating the High-fidelity Model}
\label{section:actLearn}
\vspace{-0.05in}

To build the high-fidelity model, CEAL first creates a sample pool $C_{\R{pool}}$ by selecting
configurations randomly from the workflow's
configuration space $C$. All configurations used subsequently to train 
the high-fidelity model are selected from $C_{\R{pool}}$.
As we evolve the model, we repeatedly evaluate and rank 
all configurations remaining in $C_{\R{pool}}$; thus,
we want  $|C_{\R{pool}}| \ll |C|$ to keep evaluation costs manageable.
However, we also want $C_{\R{pool}}$ to be reasonably
representative of $C$ and, in particular, to contain enough 
well-performing configurations to train a good high-fidelity model. 
Say that we want the best configuration in 
$C_{\R{pool}}$ to be in the top 1/$n$ of all 
configurations, with probability $P$.
With a pool size $p$, the chance of selecting $p$ configurations each not in 
the top 1/$n$ is less than $(1-1/n)^p$.
Thus, we want 
$p \approx -n\cdot \ln (1-P)$, because 
$P > 1-(1-1/n)^p = 1-[(1-1/n)^n]^{p/n} > 1-(1/e)^{p/n}$.
For example, if 1/$n$ = 1/500 = 0.2\% and $P$~=~98.2\%, then $p\approx$ 2000.
To establish the high-fidelity model, CEAL selects $m_{0}$
configurations at random from $C_{\R{pool}}$; 
selects the $m_{B}$ best of the configurations remaining in $C_{\R{pool}}$, based on scores from the low-fidelity model;
runs the workflow with the $m_0+m_{B}$ selected configurations; 
and uses the results as samples to train an initial high-fidelity model.
We discuss the factors influencing the choice of $m_{B}$ and $m_{0}$ below.

To evolve the model, CEAL evaluates the configurations remaining in $C_{\R{pool}}$ and 
selects the $m_{B}$ highest-scoring. %
Then, it runs the workflow with those configurations and uses those results to train the high-fidelity model further. 
It repeats these operations until
the cost budget is reached. 

The 
high-fidelity model is initially primitive and thus the low-fidelity model is a superior
choice for evaluating configurations.
As more training data are acquired, the high-fidelity model eventually
outperforms the low-fidelity model.
Thus, we monitor the capability of the high-fidelity model in evaluating configurations, 
and substitute it for
the low-fidelity model when it becomes a better choice.
Specifically, each time that we perform more runs,
we use the new data to compute the top-1, top-2, and top-3 recalls (see \S\ref{section:recallScore}) of the low- and high-fidelity models: that is, the extent to which their best 1, 2, and 3 configurations, respectively, match the best 1, 2, and 3 
as determined by experiment.
When these scores for the high-fidelity model (summed to increase stability) exceed the low-fidelity sum,
we switch to the high-fidelity model. 

As we show quantitatively when we present experimental results in \S\ref{section:evaluation},
the power of the CEAL approach derives from the fact that it selects mostly top configurations when collecting data to train the high-fidelity model.
The rationale here is this: our ultimate goal is a surrogate model that the searcher can use to find the best configurations, and for that purpose it should be highly accurate for good configurations, 
but can be less accurate for bad configurations.
Thus, we prefer to use our limited sample budget for samples collected for high-performing configurations.
(Focusing sample collection on high-performing configurations also has the useful
side effect that these samples, by definition, take less time.)

Thus, as described above, we use the low-fidelity model to bootstrap the sample selection process, 
and transition to the high-fidelity trained model as the number of samples grows.
But what if our low-fidelity model is %
biased in such a way that it never gives good scores to 
high-performing configurations? 
Then the high-fidelity model may never improve.
This concern motivates us to select, in the first phase, $m_{0}$ random configurations as well as the $m_{B}$
configurations selected with the low-fidelity model.
We discuss the sensitivity of CEAL to these hyper-parameters in \S\ref{section:hp}.

\vspace{-0.05in}

\section{CEAL Algorithm}
\label{section:algo}
\vspace{-0.05in}

The CEAL algorithm, \Algorithm{CEAL}, takes as input a data collection 
budget ($m$), expressed in terms of workflow runs for simplicity. (If a budget on real resource 
consumption is preferred, the algorithm can be adapted to monitor the
resource consumption of the workflow and its component applications.)

\MultiLines{cpntLoop}{estimate}, the low-fidelity model 
generation via component combination phase, run each component application 
$m_{\R{R}}$ times to test randomly selected configurations and build component 
models. The cost is equivalent to running the complete workflow $m_{\R{R}}$ times, 
incurring a charge of $m_{\R{R}}$ from budget $m$ (\Line{count}). If a component
application has been tested earlier in other workflows, some configuration-performance
data $D_j^{\R{hist}}$ can be reused to further improve component model quality
(\Line{data}). In the case, $m_{\R{R}}$ should be close to 0.
Otherwise, $m_{\R{R}}$ is generally set to be from $m \cdot 20\%$ to $m \cdot 70\%$.

\begin{algorithm}[!t]
\footnotesize
\caption{\small %
\B{C}omponent-based \B{E}nsemble \B{A}ctive \B{L}earning
{
\footnotesize
\newline \B{Inputs:}
    Workflow runs budget $\bm{m}$;
    budget used to run component applications $\bm{m_{\B{\R{R}}}}$;
    for each component application $1 \le j \le J$, 
    configuration space $\bm{C}_j$
    and historical measured configuration-performance samples $\bm{D}_j^{\B{\R{hist}}}$; 
    sample pool $\bm{C}_{\B{\R{pool}}}$;
    number of initial random samples $\bm{m_0}$; 
    number of iterations $\bm{I}$.
\newline \B{Output:} High-fidelity workflow model $\bm{M_{\R{H}}}$.
}
}
\label{algorithm:CEAL}
    \begin{algorithmic}[1]
        \For {$j \in \{1, 2, \cdots, J\}$}    \label{line:cpntLoop}
            \State Randomly select $m_{\R{R}}$ configurations from $C_j$ as $C_j^{\R{meas}}$;    \label{line:srcRandSlct}
            \State Run $j$\textsuperscript{th} component with 
            configurations in $C_j^{\R{meas}}$, giving $D_j^{\R{meas}}$;    \label{line:srcMeas}
            \State $D_j^{\R{meas}} \leftarrow D_j^{\R{meas}} \cup D_j^{\R{hist}}$;    \label{line:data}
            \State Train component model $M_j^{\R{cpnt}}$ with $D_j^{\R{meas}}$ for the $j$\textsuperscript{th} application; \label{line:srcTrain}
        \EndFor
        \State Generate low-fidelity model $M_{\R{L}}$ from 
        component models and combination function (see \SL{combo});    \label{line:estimate}
        \State Move $m_0$ randomly selected configurations from $C_{\R{pool}}$ to form $C_{\R{meas}}$;    \label{line:tgtRandSlct}
        \State $m_B \leftarrow (m - m_0 - m_{\R{R}}) / I$;    \label{line:count}
        \State Score all configurations in $C_{\R{pool}}$ with $M_{\R{L}}$;    \label{line:m0_evaluate}
        \State Move top $m_B$ configurations from $C_{\R{pool}}$ into $C_{\R{meas}}$;    \label{line:topSlct}
        \State $M \leftarrow M_{\R{L}}$; \ \ \ \ // Set the model used for evaluating configurations
        \State $M_{\R{H}} \leftarrow$ ML model (e.g., boosted 
        tree~\cite{Chen:XGBoostScalTreeBoosSys:KDD16}); 
        // Init high-fidelity model  \label{line:initMdl}
        \For {$i \in \{1, \cdots, I\}$}    \label{line:beginRefine}
            \State Run workflow for all configurations in $C_{\R{meas}}$, giving $D_{\R{meas}}$;    \label{line:meas}
            \If{$M = M_{\R{L}}$}\ \ \ \ // Begin model switch detection     \label{line:detStart}
               \State // Score models: recalls for top 1, 2, 3 configurations
                \State $S_{\R{H}} = \sum_{i=1,2,3} S_r(i, C_{\R{meas}}, M_H, D_{\R{meas}}) $; \ \ // See \S\ref{section:recallScore}
                \State $S_{\R{L}} = \sum_{i=1,2,3} S_r(i, C_{\R{meas}}, M_L, D_{\R{meas}}) $; \ \ \ // See \S\ref{section:recallScore}
                \State \textbf{if} $S_{\R{H}} \ge S_{\R{L}}\ \textbf{then}\  M \leftarrow M_{\R{H}}$ \textbf{endif}; \label{line:switch}
            \EndIf \ \ \ \ \ // End model switch detection    \label{line:detEnd}
            \State Use $D_{\R{meas}}$ to train/refine $M_{\R{H}}$, and update $M$ if it switched to $M_{\R{H}}$;    \label{line:tgtTrain}
            \State Use $M$ to evaluate the configurations in $C_{\R{pool}}$;
            \State Move the top $m_B$ configurations from $C_{\R{pool}}$ to form $C_{\R{meas}}$ again; \label{line:tgtPred}
        \EndFor    \label{line:endRefine}
        \State \Return $M_{\R{H}}$.    \label{line:mdlGen}
    \end{algorithmic}
\end{algorithm}

The second phase, high-fidelity model generation via dynamic ensemble active learning,
is \MultiLines{tgtRandSlct}{mdlGen}. First, $m_0$ random training samples 
are selected (\Line{tgtRandSlct}) to characterize the overall performance distribution
of the workflow over all configurations.
In general, $m_0$ is set to be from $m \cdot 5\%$ to $m \cdot 45\%$, 
depending on %
workflow structure and optimization objective. 
Sensitivity studies reported in \SL{hp} show that CEAL is insensitive to the 
values used for $m_{\R{R}}$ and $m_0$ in a large range. 
We recommend $m_0 \approx 25\% \cdot m$ when $|D_j^{\R{hist}}| \gg m$ ($j$=1,...,$J$) 
and thus $m_{\R{R}}$=0, and $m_0 \approx 15\% \cdot m$ if no historical measurements 
are available (i.e., $|D_j^{\R{hist}}|$=0).
Then, top configurations are selected (\TwoLines{topSlct}{tgtPred}) based on the 
evaluation with model $M$. (\MultiLines{detStart}{detEnd} handle the switching 
from low-fidelity to high-fidelity model.)
The high-fidelity model is retrained repeatedly as more training data are acquired
(\MultiLines{beginRefine}{endRefine}), and returned as the output of 
the algorithm.

\vspace{-0.05in}
\section{Experimental Evaluation}
\label{section:evaluation}

We describe %
our benchmarks (\SL{benchmarks}), evaluation metrics (\SL{metrics}), 
and comparison targets (\SL{comparison}). 
Then, we evaluate the performance of CEAL and other algorithms in a 
general auto-tuning scenario without historical measurements, 
and investigate reasons for CEAL's superiority (\SL{noHist}). 
We also consider %
optimization 
with historical component
measurements, and compare CEAL with an algorithm that incorporates 
component performance by training a ML model (\SL{evalHist}). 
Finally, we study CEAL's sensitivity to hyper-parameter values (\SL{paraSens}).

\vspace{-0.1in}
\subsection{Experimental Setup}
\label{section:benchmarks}
\vspace{-0.05in}

We conducted experiments on a 600-node cluster with Intel Omni-Path Fabric Interconnect. 
Each node has two 18-core 2.10GHz Intel Broadwell Xeon E5-2695 v4 processors 
with hyperthreading disabled and 128 GB DDR4 SDRAM.
We ran each workflow with exclusive access to node 
resources, on allocation sizes up to 32 compute nodes.
We used three in-situ workflows,
namely \B{LV}, \B{HS}, and \B{GP}, in our experiments:

    \B{LV} couples two components:
    the LAMMPS~\cite{LAMMPS} molecular dynamics
    simulator and Voro++~\cite{Voropp}, a Voronoi tesselator.
    LV involves full-featured, realistic applications coupled via ADIOS.
    The sample run used here simulates \num{16000} atoms and streams position and
    velocity data into the tesselator for analysis and visualization.
    This application is a model for many cases in particle simulation and
    visualization.

    \B{HS} also couples two components:
    a Heat Transfer~\cite{HeatTransfer} simulation
    with an analysis application, Stage Write.
    Heat Transfer is a mini-application that runs the heat equation over the grid of a given size and forwards simulation
    state over ADIOS to Stage Write, which produces output in the file system.
    This application is a model for many cases in numerical PDE calculations and
    I/O buffering and forwarding.

    \B{GP} couples four components:
    Gray-Scott reaction-diffusion simulation; an analysis
    application, PDF calculator, applied to the Gray-Scott output;
    a visualization application, G-Plot, also applied to Gray-Scott output;
    and a second visualization application, P-Plot, applied to PDF output.
    GP involves applications of intermediate complexity, also coupled via ADIOS.
    Two component applications, Gray-Scott and PDF calculator
    are configurable, but G-Plot and P-Plot are not.
    This application is a model for many cases in chemical reaction dynamics and
    more complex multi-purpose analysis workflows.

\begin{table}
\begin{center}
\caption{Parameter spaces for our three target workflows}
\vspace{-3ex}
\scriptsize
\begin{tabular}{|c|c|l|l|}
    \hline
    \B{Workflow} & \B{Application} & \B{Parameter} & \B{Options} \\
    \hline
    & & \# processes & 2, 3, $\cdots$, \num{1085} \\
    \cline{3-4}
    & LAMMPS & \# processes per node & 1, 2, $\cdots$, 35\\
    \cline{3-4}
    LV & & \# threads per process & 1, 2, 3, 4 \\
    \cline{3-4}
    & & \# steps in an IO interval & 50, 100, $\cdots$, 400 \\
    \cline{2-4}
    & & \# processes & 2, 3, $\cdots$, \num{1085} \\
    \cline{3-4}
    & Voro++ & \# processes per node & 1, 2, $\cdots$, 35 \\
    \cline{3-4}
    & & \# threads per process & 1, 2, 3, 4 \\
    \hline
    & & \# processes in X & 2, 3, $\cdots$, 32 \\
    \cline{3-4}
    & Heat & \# processes in Y & 2, 3, $\cdots$, 32 \\
    \cline{3-4}
    HS & transfer & \# processes per node & 1, 2, $\cdots$, 35 \\
    \cline{3-4}
    & & \# IO writes & 4, 8, $\cdots$, 32 \\
    \cline{3-4}
    & & buffer size (MB) & 1, 2, $\cdots$, 40 \\
    \cline{2-4}
    & Stage & \# processes & 2, 3, $\cdots$, \num{1085} \\
    \cline{3-4}
    & write & \# processes per node & 1, 2, $\cdots$, 35 \\
    \hline
    & Gray- & \# processes & 2, 3, $\cdots$, 1085 \\
    \cline{3-4}
    GP & Scott & \# processes per node & 1, 2, $\cdots$, 35 \\
    \cline{2-4}
    & PDF & \# processes & 1, 2, $\cdots$, 512 \\
    \cline{3-4}
    & calculate & \# processes per node & 1, 2, $\cdots$, 35 \\
    \cline{2-4}
    & Gray plot & \# processes & 1 \\
    \cline{2-4}
    & PDF plot & \# processes & 1 \\
    \hline
\end{tabular}
\label{table:parameter}
\vspace{-4ex}
\end{center}
\end{table}

Application configuration options, shown in~\Table{parameter}, form a total of
2.3 $\times$ 10$^{10}$ possible configurations for LV (LAMMPS: 6.1 
$\times$ 10$^{5}$;  Voro++: 7.6 $\times$ 10$^{4}$),
5.1 $\times$ 10$^{10}$ for HS (Heat Transfer: 5.4 
$\times$ 10$^{6}$;  Stage Write: 1.9 $\times$ 10$^{4}$),
and 8.5 $\times$ 10$^{7}$ for GP (Gray-Scott:
1.9 $\times$ 10$^{4}$; PDF calculator: 9.0 $\times$ 10$^{3}$).
We obtained expert-recommended configurations for each.

In order to compare expert-recommended vs.\ good configurations,
we generated for each workflow, as $C_{\R{pool}}$, 2000 configurations
of randomly selected parameter values.
Then, for each such configuration, 
we launched all workflow components at once and recorded each component's end-to-end wall-clock time. 
We then used the longest component execution time as the configuration's \emph{execution time},
and the product of execution time, number of computing nodes used, and
number of cores per node as the configuration's \emph{computer time}.
We list in \Table{perfConf} the expert-recommended and best configurations 
for each workflow and optimization objective, and their achieved performance.
The expert recommendations only do well for GP.
(Since the unconfigurable G-Plot is the bottleneck in GP,
many GP configurations have similar execution times, close to that
of G-Plot alone, 97.0 seconds.)

\begin{table}
\begin{center}
\caption{Configurations and performance of benchmarks}
\vspace{-3ex}
\scriptsize
\begin{tabular}{|c|c|c|c|c|}
    \hline
    \B{Wf.} & \B{Objective} & \B{Option} & \B{Performance} & \B{Configuration} \\
    \hline
    & Exec. time & Best & 27.2 secs & (430, 23, 1, 300, 88, 10, 4) \\  %
    \cline{3-5}
    LV & & Expert & 36.8 secs & (288, 18, 2, 400, 288, 18, 2) \\  %
    \cline{2-5}
    & Comp. time & Best & 3.36 core-hrs & (175, 35, 2, 400, 38, 29, 3) \\  %
    \cline{3-5}
    & & Expert & 4.15 core-hrs & (18, 18, 2, 400, 18, 18, 2) \\  %
    \hline
    & Exec. time & Best & 6.02 secs & (13, 17, 14, 4, 29, 19, 3) \\
    \cline{3-5}
    HS & & Expert & 28.0 secs & (32, 17, 34, 4, 20, 560, 35) \\  %
    \cline{2-5}
    & Comp. time & Best & 0.517 core-hrs & (5, 25, 35, 4, 3, 5, 3) \\ %
    \cline{3-5}
    & & Expert & 0.894 core-hrs & (8, 4, 32, 4, 20, 35, 35) \\  %
    \hline
    & Exec. time & Best & 98.7 secs & (175, 13, 24, 23, 1, 1, 1) \\  %
    \cline{3-5}
    GP & & Expert & 102 secs & (525, 35, 525, 35, 1, 1) \\
    \cline{2-5}
    & Comp. time & Best & 6.95 core-hrs & (66, 34, 41, 22, 1, 1) \\  %
    \cline{3-5}
    & & Expert & 5.85 core-hrs & (35, 35, 35, 35, 1, 1) \\
    \hline
\end{tabular}
\label{table:perfConf}
\vspace{-5ex}
\end{center}
\end{table}

We also measured the execution and computer times of 500 configurations 
randomly selected from the parameter space of each configurable component application,
and used these samples as component measurements, from which CEAL may select 
training samples for component models.

\vspace{-0.05in}
\subsection{Evaluation Metrics}
\label{section:metrics}
\vspace{-0.05in}

We use three metrics to evaluate auto-tuning algorithms.

\vspace{-0.05in}
\subsubsection{Performance of Best Predicted Configuration} \label{section:actPerf}
As our goal is to optimize the execution time and computer time of 
in-situ workflows, the execution and computer time achieved by the 
best configuration predicted for a workflow is an important evaluation metric.

\vspace{-0.05in}
\subsubsection{Robustness in Finding Top Configurations} \label{section:recallScore}
We use the \B{recall score} to measure the error tolerance of an autotuning algorithm
in predicting top configurations.
As defined by Marathe~\etal~\cite{Marathe:PerfMdlResConsTransLearn:SC17},
recall score is, for a value $n$, the percentage of configurations as predicted by the
algorithm that are actually within the top $n$ configurations.
Given a set of configurations \textbf{c} for which we have workflow performance data $\textbf{D}_{\textbf{c}}$,
a model \textbf{M} for scoring configurations,
and a function top$(n, s)$ for selecting the top $n$ entries from a set of scored configurations $s$,
the recall score for $n$ is:
\vspace{-1ex}
\begin{equation}
S_r(n, \textbf{c}, \textbf{M}, \textbf{D}_{\textbf{c}}) = | \textrm{top}(n, \textbf{M}(\textbf{c})) \cap \textrm{top}(n, \textbf{D}_{\textbf{c}})|/n \times 100\%.
\end{equation}
A higher recall score indicates a more robust auto-tuning algorithm;
in general, $S_r(i)$ is more important than $S_r(j)$ ($i < j$).
For $n=1$, the recall score also represents the probability of finding the
best-performing configuration.

\vspace{-0.05in}
\subsubsection{Practicality in Performance Optimization} \label{section:netGain}

Since the data collection cost is considerable and unignorable for empirical 
auto-tuning algorithms, we monitor the least number of workflow runs required 
to pay off the auto-tuning cost, and use that metric to evaluate the practicality 
of auto-tuning algorithms. 
The \B{least number of uses} is $N=c/\Delta p$.
Here, $\Delta p$ is the actual improvement per workflow execution
in the optimization objective (execution time or computer time reduction) achieved by
the auto-tuning algorithm relative to an expert recommendation, and
$c$ is the training data collection cost incurred in achieving the
performance optimization objective, i.e., the sum of the workflow's execution times or 
computer times over all training samples.

\vspace{-0.05in}
\subsection{Comparison Targets}
\label{section:comparison}
\vspace{-0.05in}

\begin{figure*}
\begin{center}
  \begin{minipage}[t]{1\linewidth}
    \begin{subfigure}[t]{.385\linewidth}
     \includegraphics[height=3.05cm, width=1\linewidth]{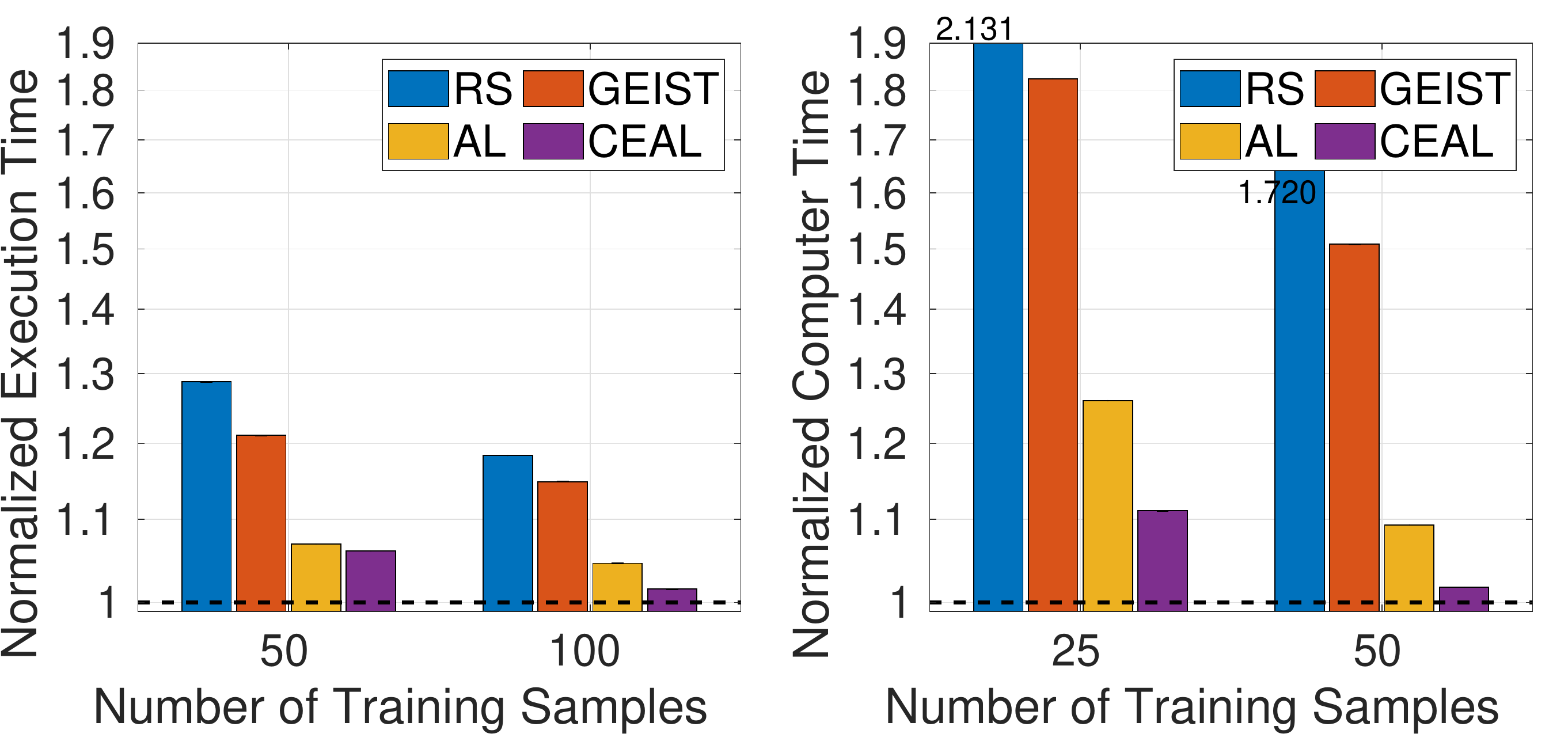}
     
           \vspace{-1ex}
           
      \caption{\it LV}
      \label{figure:topPerf_lv}
    \end{subfigure}
    \hspace{0.005\linewidth}
    \begin{subfigure}[t]{.385\linewidth}
      \includegraphics[height=3.05cm, width=1\linewidth]{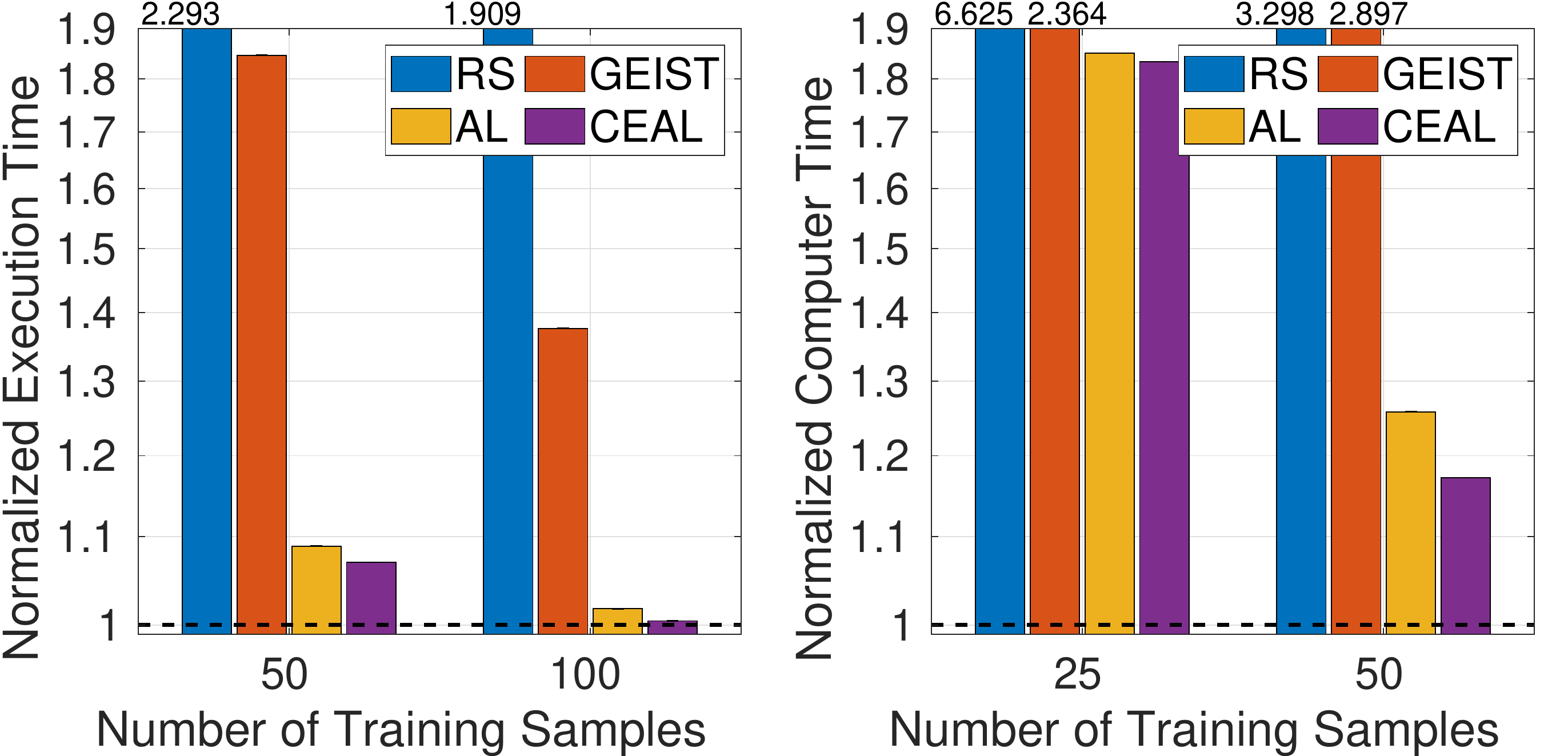}
      
            \vspace{-1ex}
            
      \caption{\it HS}
      \label{figure:topPerf_hs}
    \end{subfigure}
    \hspace{0.005\linewidth}
    \begin{subfigure}[t]{.193\linewidth}
      \includegraphics[height=3.05cm, width=1\linewidth]{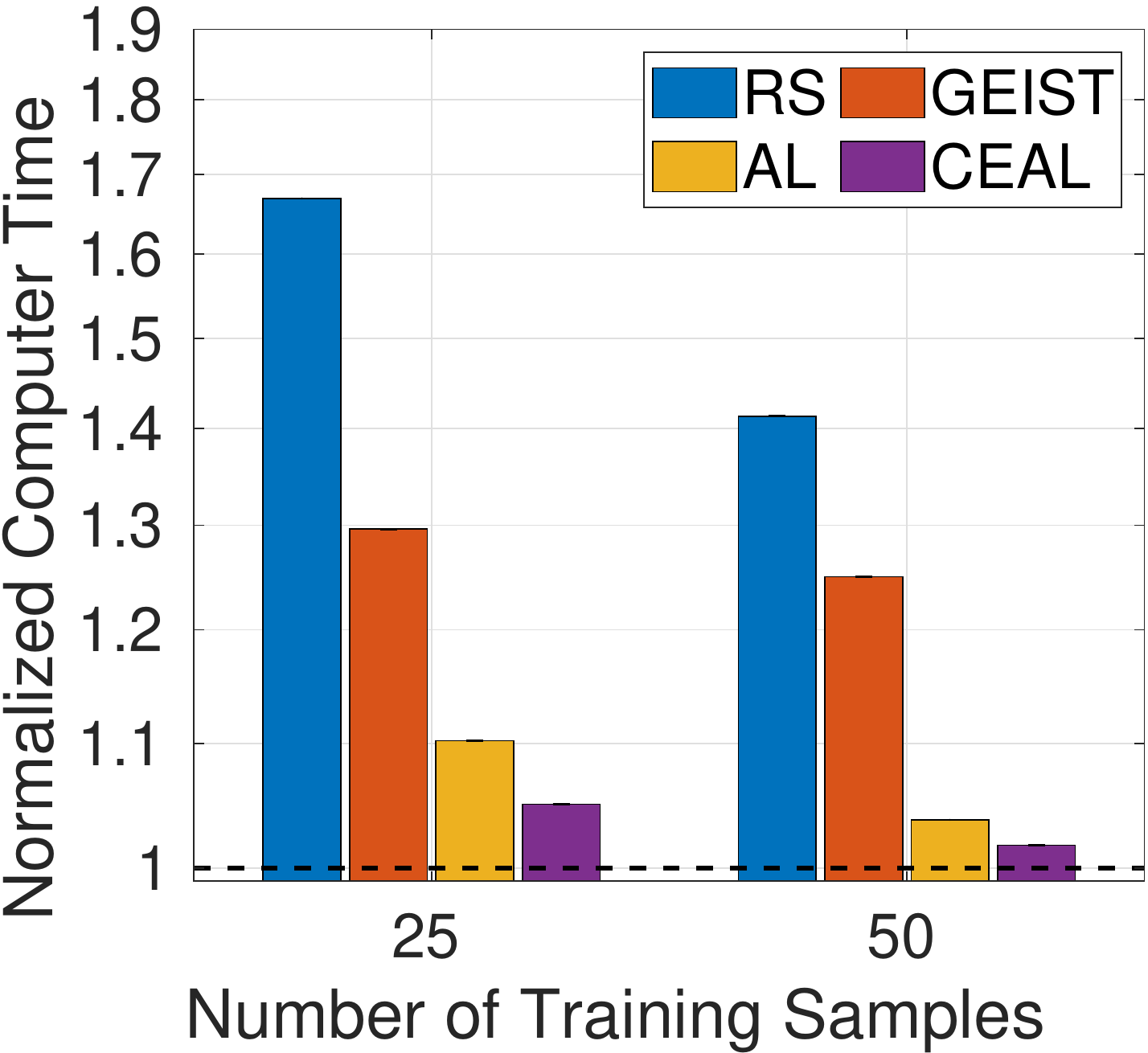}
      
      \vspace{-1ex}
      
      \caption{\it GP}
      \label{figure:topPerf_gp}
    \end{subfigure}
    \vspace{-3ex}
      \caption{The best configuration auto-tuned w/o historical measurements (dashed lines: the best configuration in the test set)}
    \label{figure:topPerf}
  \end{minipage}
\end{center}
\vspace{-2ex}
\end{figure*}

\begin{figure*}[!t]
\begin{center}
  \begin{minipage}[t]{.32\linewidth}
    \includegraphics[height=3.05cm, width=1\linewidth]{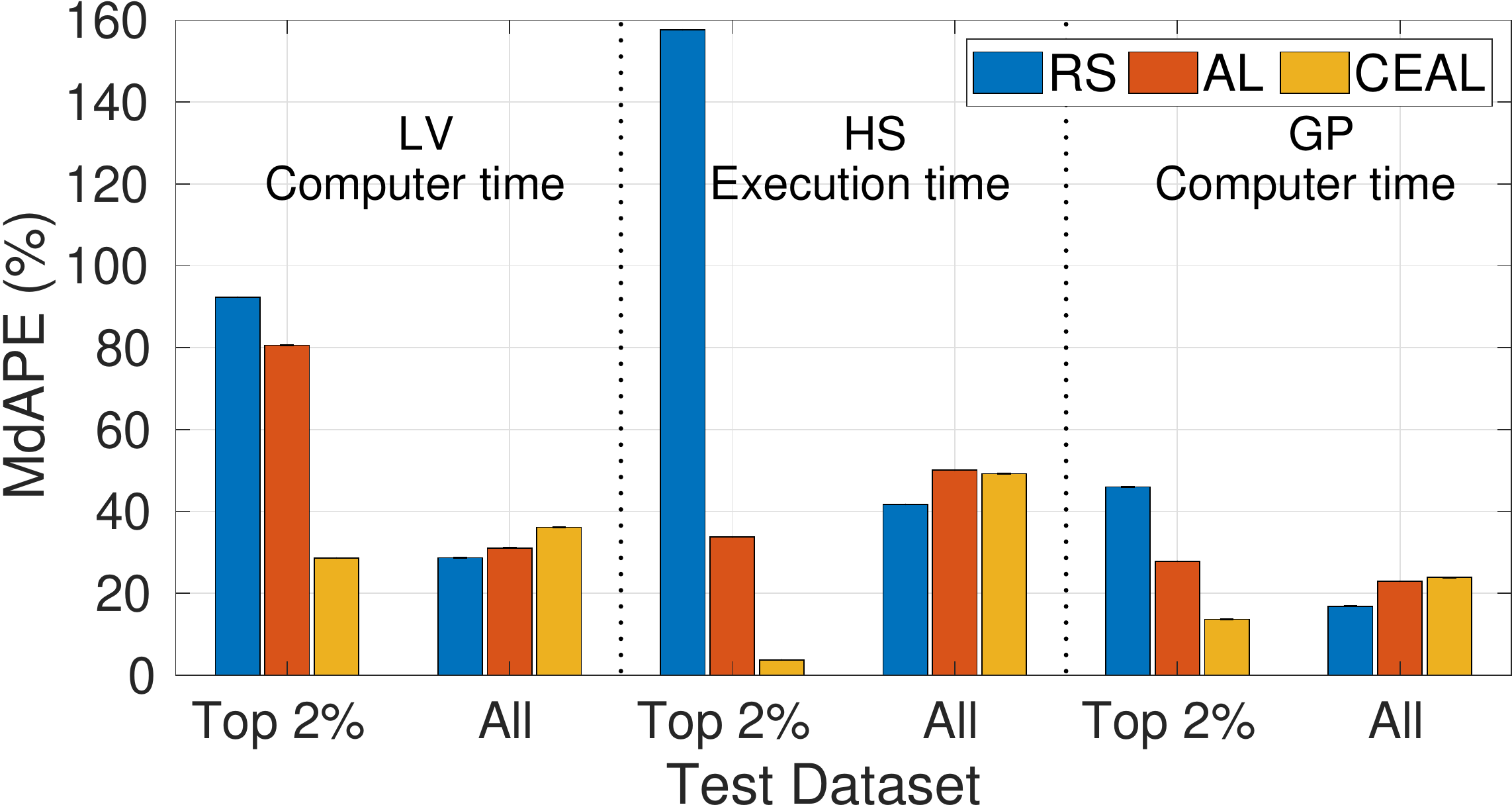}
    \vspace{-5ex}
    \caption{Prediction accuracy of models in autotuning w/o historical measurements}
    \label{figure:mdape}
  \end{minipage}
  \hspace{0.005\linewidth}
  \begin{minipage}[t]{.65\linewidth}
    \begin{subfigure}[t]{.485\linewidth}
      \includegraphics[height=3.05cm, width=1\linewidth]{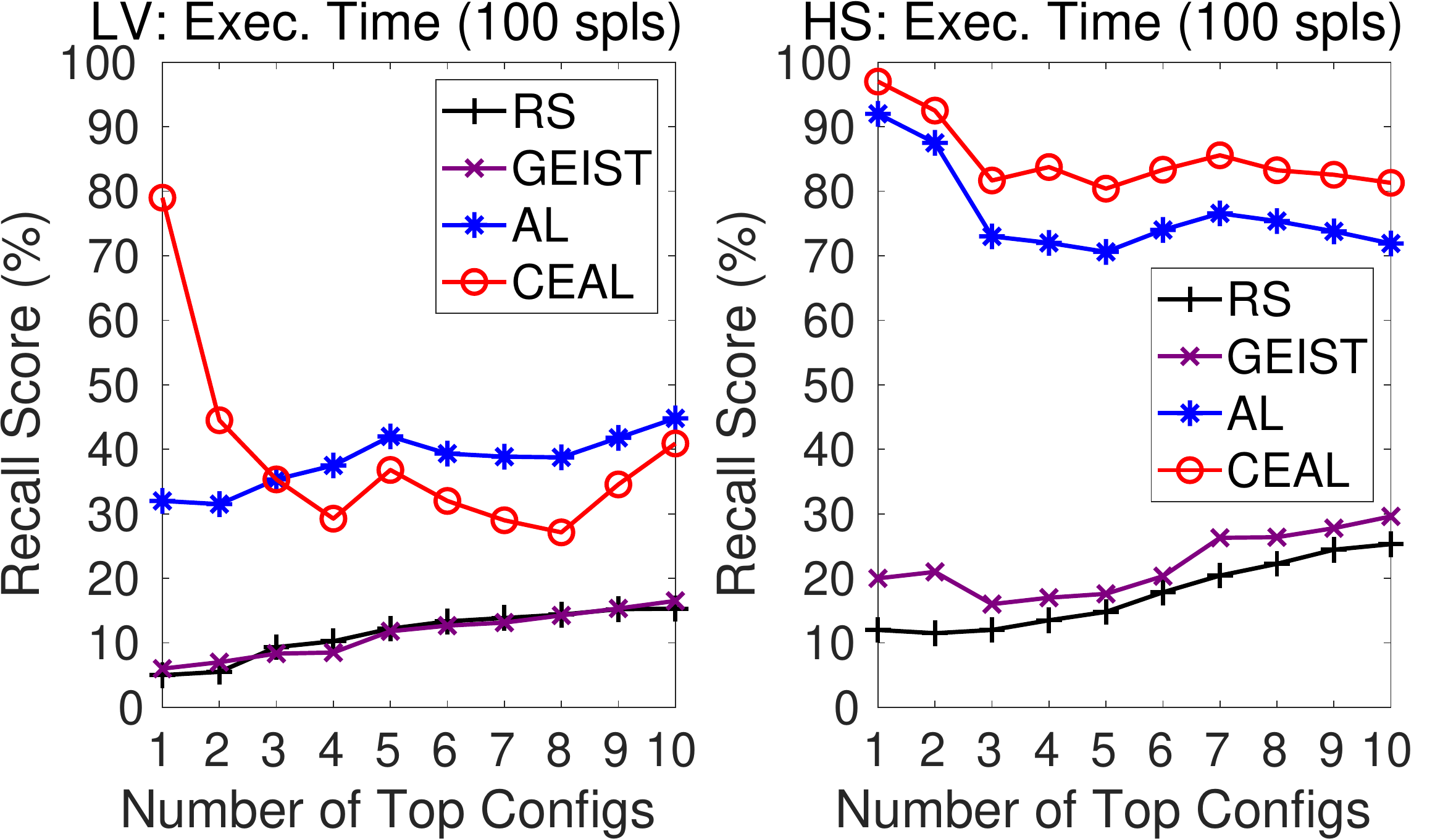}
      
            \vspace{-1ex}
            
      \caption{\it Optimizing execution time of LV and HS}
      \label{figure:rs_exec}
    \end{subfigure}
    \hspace{0.005\linewidth}
    \begin{subfigure}[t]{.485\linewidth}
      \includegraphics[height=3.05cm, width=1\linewidth]{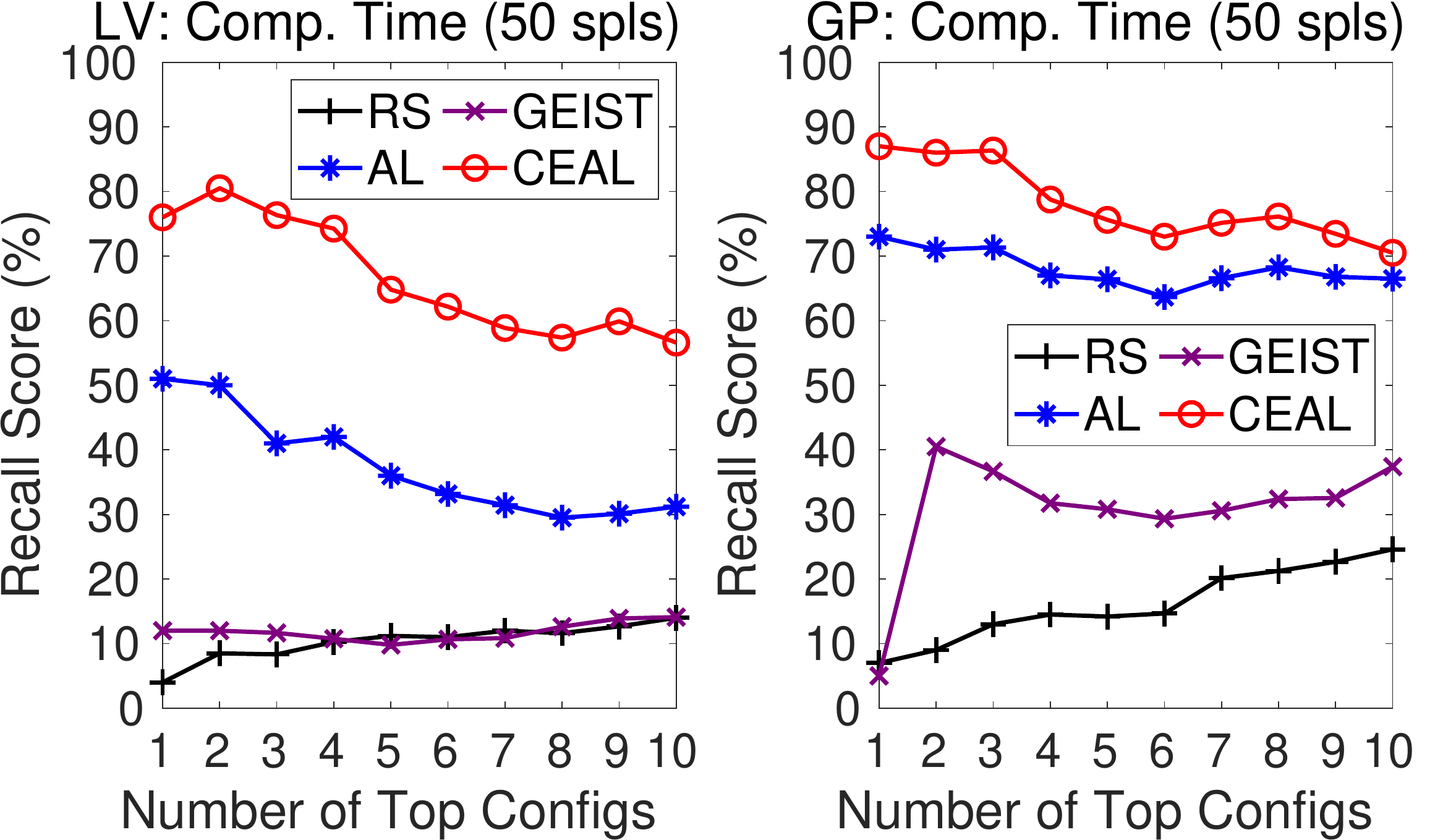}
      
            \vspace{-1ex}
            
      \caption{\it Optimizing computer time of LV and GP}
      \label{figure:rs_comp}
    \end{subfigure}
    \vspace{-2.5ex}
    \caption{Robustness of autotuning w/o historical measurements}
    \label{figure:rs}
  \end{minipage}
\end{center}
\vspace{-3ex}
\end{figure*}

Since there exist few auto-tuning algorithms customized for in-situ workflows,
we compare CEAL with three auto-tuning algorithms
for general HPC applications. %
    \B{RS} selects training data by 
    random sampling.
    \B{AL} is a typical AL algorithm that iteratively selects a batch 
    of the best configurations predicted by gradually refined models as training 
    samples~\cite{Mametjanov:AutotunFPGAParaPerfPow:FCCM15,
    Behzad:OptIOPerfHPCAppAutotun:TOPC19}.
    \B{GEIST}, a state-of-the-art AL-based auto-tuning algorithm for finding
    performance-optimizing configurations~\cite{Thiagarajan:BootParaSpacExplFastTun:ICS18}
    is guided by a parameter graph to choose training samples with the
    highest possibility of being optimal (defined as in top 5\% configurations)
    in each iteration.
We also report in \S\ref{sec:alph} on comparisons with a variant of CEAL, \B{ALpH} (introduced in \SL{combo})
that uses learning to combine component models.

In all algorithms, we use the xgboost.XGBRegressor
implementation of 
extreme gradient boosting regression as the original ML model.
We adjust GEIST, AL, ALpH, and CEAL hyperparameters with and without
historical measurements, and $I$, $m_0$, and $m_{\R{R}}$ in CEAL, and 
select the best settings for each algorithm. %
Hyperparameter optimization methods~\cite{Xia:BoosDeciTreeApprBayeHypeOptiCredScor:ESA17} are beyond the scope of this paper.
In all experiments, we run each algorithm 100 times and report averages.

\vspace{-0.05in}
\subsection{Autotuning without Historical Measurements}
\vspace{-0.05in}
\label{section:noHist}

\begin{figure*}[!t]
\begin{center}
  \begin{minipage}[t]{.3\linewidth}
    \includegraphics[height=3cm, width=1\linewidth]{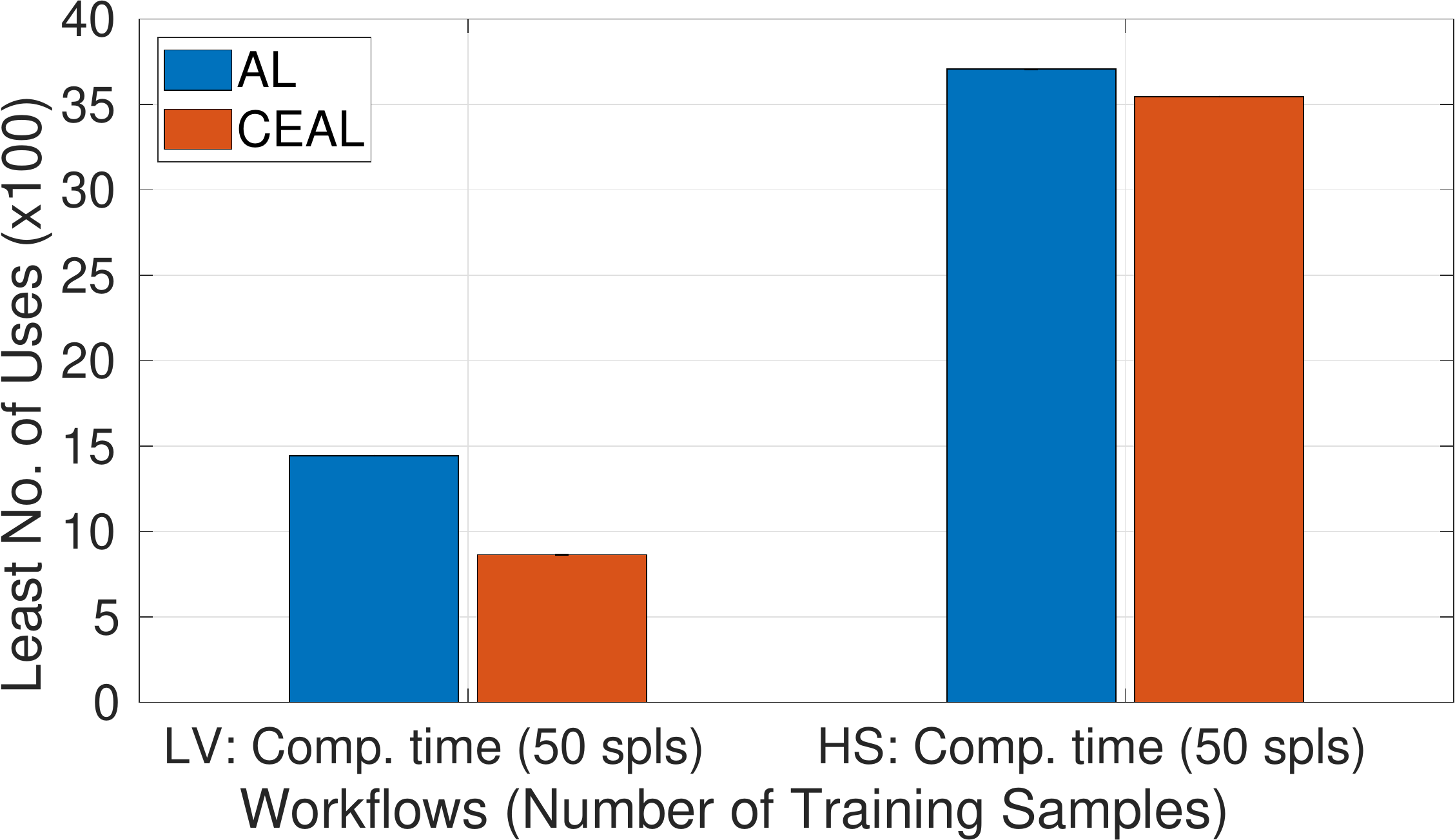}
    
    \vspace{-2ex}
    
    \caption{Practicality of autotuning w/o historical measurements}
    \label{figure:numUse_comp}
  \end{minipage}
  \hspace{0.005\linewidth}
  \begin{minipage}[t]{.67\linewidth}
    \begin{subfigure}[t]{.485\linewidth}
      \includegraphics[height=3.05cm, width=1\linewidth]{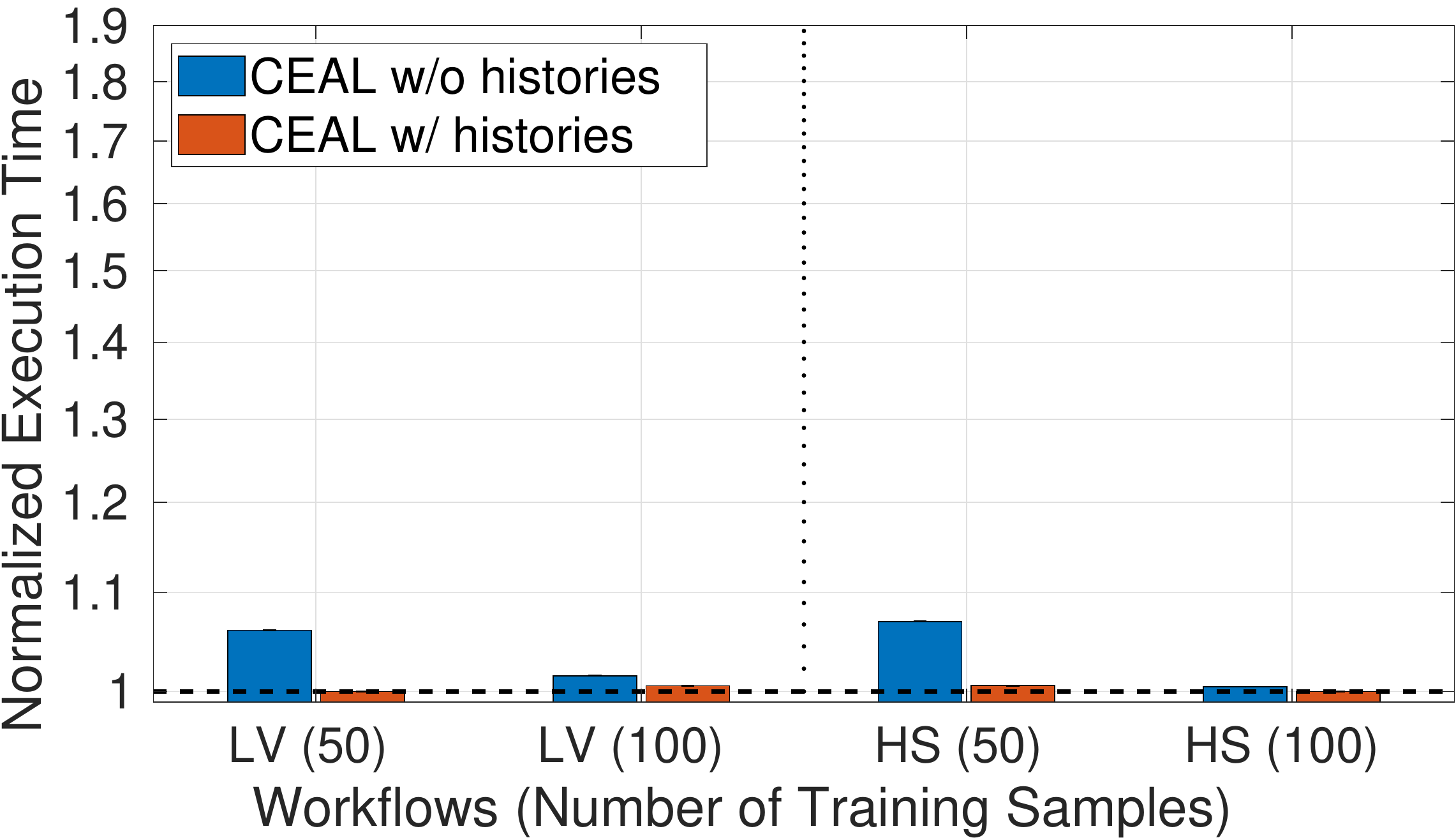}
      
      \vspace{-1ex}
      
      \caption{\it Execution time of predicted best conf.}
      \label{figure:topPerf_ceal_exec}
    \end{subfigure}
    \hspace{0.005\linewidth}
    \begin{subfigure}[t]{.485\linewidth}
      \includegraphics[height=3.05cm, width=1\linewidth]{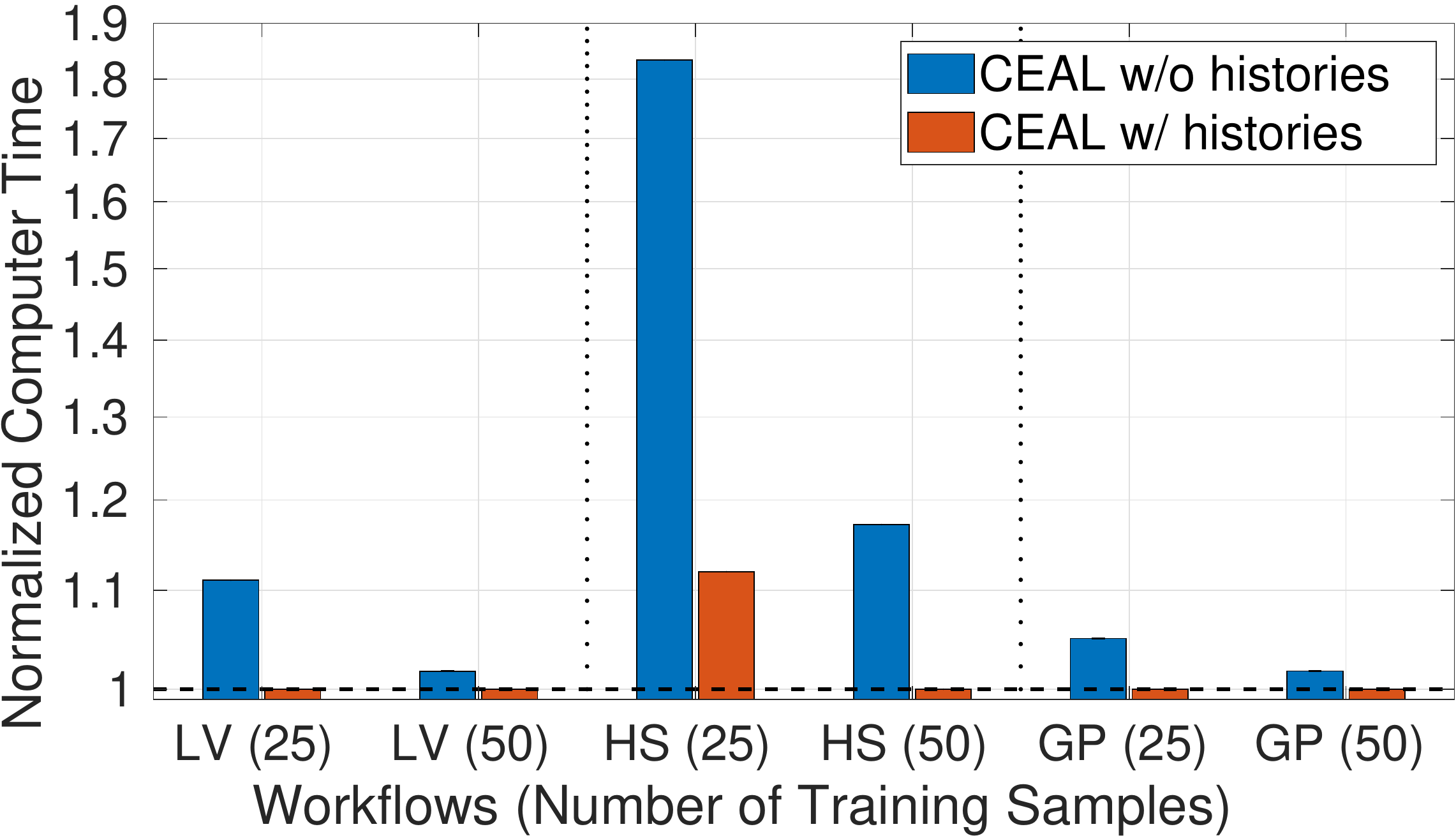}
      
            \vspace{-1ex}
            
      \caption{\it Computer time of predicted best conf.}
      \label{figure:topPerf_ceal_comp}
    \end{subfigure}
    \vspace{-2.5ex}
      \caption{Effect of historical measurements on CEAL (dashed lines: best test config.)}
    \label{figure:topPerf_ceal}
  \end{minipage}
\end{center}
\vspace{-3ex}
\end{figure*}

We first examine the overall performance of our auto-tuner in the absence of 
historical measurements. 
We compare the actual performance of workflows auto-tuned by CEAL and others (\SL{NHperf}), 
and explain CEAL's superiority by experimentally validating its design principle (\SL{actReason}). 
Also, we investigate CEAL's robustness and practicality (\SL{NHrobust} and \SL{NHpractice}).
(When comparing costs, we consider the cost of running an in-situ workflow as being comparable to the 
total cost of running all of its component applications separately.)

\subsubsection{Actual Performance of Auto-tuned Workflows}
\label{section:NHperf}
We measured the actual execution and computer time of the best configurations of
LV, HS, and GP predicted by RS, GEIST, AL, and CEAL,
and plot normalized values in~\Figure{topPerf}, with the performance of the 
best configuration in the test set shown as ``1'' (the same for \TwoFigures{topPerf_ceal}{topPerf_hist}).
We test CEAL with different numbers $m$ of training samples by doubling $m$ from 
25 until the auto-tuned performance of LV is at most 5\% worse than the best. 
We show here results for the largest two values of $m$ tested: 100 and 
50 for execution time and 50 and 25 for computer time. 
For consistency, we also select the same $m$ for all workflows in all experiments.
\Figure{topPerf} shows that the execution and computer times achieved by CEAL 
are always better than by RS, GEIST, and AL.
For example, CEAL improves both execution and computer time
by 14--72\% relative to RS and 12--60\% relative to GEIST---and LV computer time relative to LV by 11.9\%
and 6.9\% with 25 and 50 training samples, respectively.
CEAL outperforms AL, because performance models trained with the same
number of training samples are much more accurate for component applications
than in-situ workflows,
and our method of determining workflow performance provides relatively
accurate configuration ranking over top configurations.

\vspace{-0.05in}
\subsubsection{Why CEAL Outperforms RS and AL}
\label{section:actReason}
The absolute percentage error (APE) of a sample $i$ is
$|(y_i - y'_i) / y_i|$, 
where $y_i$ is actual performance and $y'_i$ is predicted performance.
The median APE (MdAPE) for a set of samples is a commonly used measure of model prediction quality.
To further understand why CEAL beats AL and RS, we plot in \Figure{mdape} the
MdAPEs of models generated by RS, AL, and CEAL
when used to predict performance over all, and the top 2\%, of test dataset configurations.
CEAL MdAPEs are much less than those of RS and AL
for the top 2\% of configurations;
as a result, CEAL outperforms RS and AL,
even though its MdAPEs are comparable to, or a little higher than, those
of RS and AL over \emph{all} configurations.
This result verifies our intuition that picking higher-performance configurations
can improve prediction accuracy for top configurations,
and thus make best use of the few training samples allotted.

\vspace{-0.05in}
\subsubsection{Robustness of Auto-tuning Algorithms}
\label{section:NHrobust}
We use the recall scores (\SL{recallScore}) of the top $n$ ($n=1, \cdots, 10$)
configurations to evaluate the robustness of RS, GEIST, AL, and CEAL in
auto-tuning our workflows %
for execution time and computer time.
We see in \Figure{rs} 
that CEAL is more robust than RS, GEIST, and AL in most cases.
For the top-one configuration recall score (the most important performance measure),
CEAL achieves 76\% (or 79\%) when optimizing the computer (or execution) time 
of LV with 50 training samples,
as against 4\% (or 5\%), 12\% (or 6\%), and 51\% (or 32\%) for RS, GEIST, and AL.
For instance, although CEAL's recall score is lower than that of AL for the top 
4 to 10 configurations for LV, it still obtains a better execution time.

\vspace{-0.05in}
\subsubsection{Practicality of Auto-tuning Algorithms}
\label{section:NHpractice}
We examine the practicality of the four auto-tuning algorithms in optimizing
the computer time of LV and HS.
Since the computer time of LV/HS achieved by RS and GEIST with only 25 and 50
training samples is worse than that at the expert-recommended configuration,
the practicality of RS and GEIST is limited.
Then, we focus on auto-tuning the computer time of LV and HS by AL and CEAL 
with 50 training samples,
and plot the least number of runs (\SL{netGain}) in~\Figure{numUse_comp},
which reveals that CEAL is superior to AL in terms of practicality.
In particular, LV is worth auto-tuning by CEAL if it is expected to run 864 times,
40\% less than the 1444 times required for AL.
We attribute CEAL's superiority to its more accurate selection of
training samples that take less computer time, as boosted by the combined 
low-fidelity model.

\vspace{-0.05in}
\subsection{With Historical Component Measurements}
\label{section:evalHist}
\vspace{-0.05in}

\begin{figure*}[!t]
\begin{center}
  \begin{minipage}[t]{.65\linewidth}
    \begin{subfigure}[t]{.485\linewidth}
      \includegraphics[height=3.05cm, width=1\linewidth]{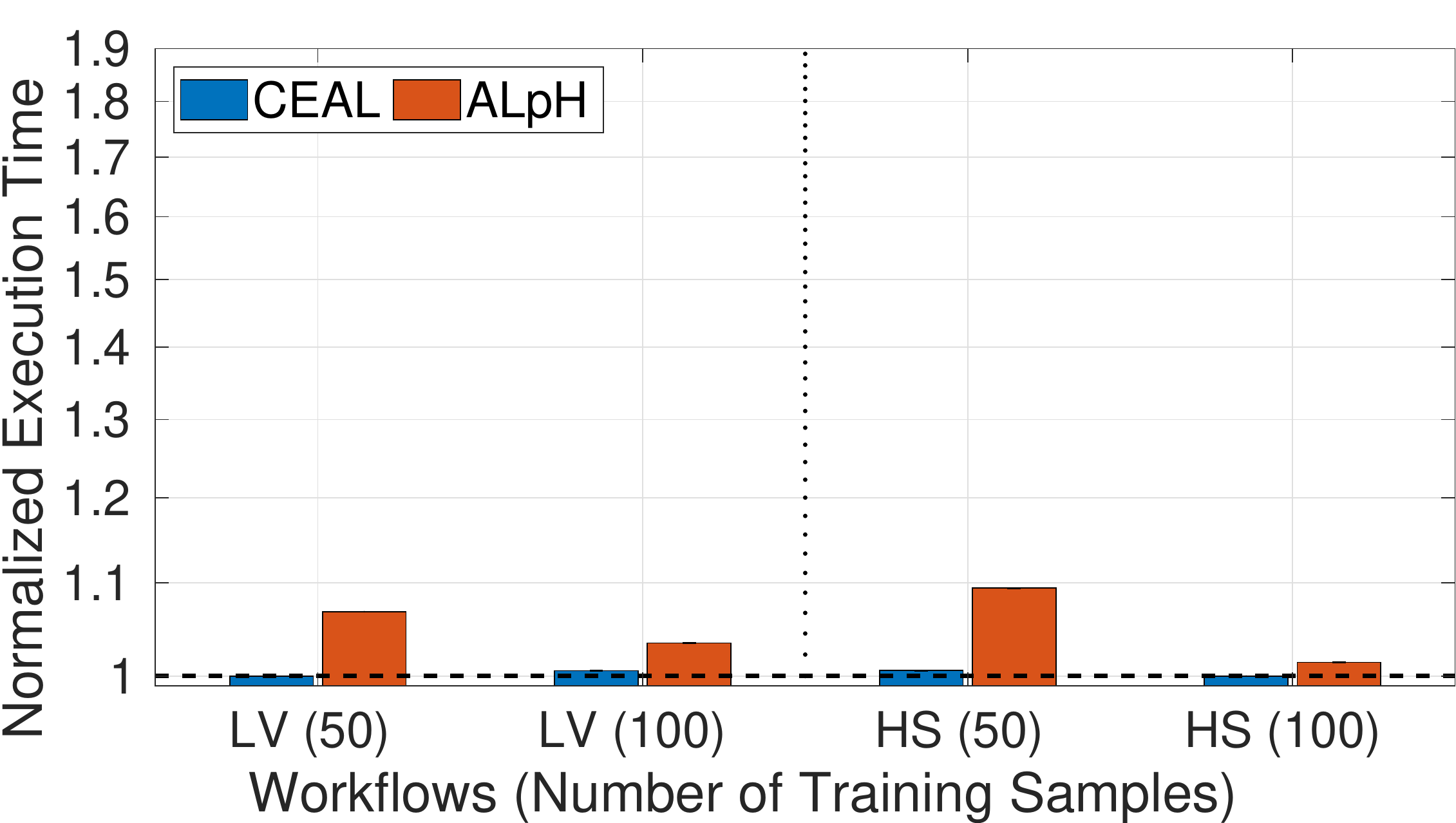}
      
            \vspace{-1ex}
            
      \caption{\it Optimizing execution time}
      \label{figure:topPerf_hist_exec}
    \end{subfigure}
    \hspace{0.005\linewidth}
    \begin{subfigure}[t]{.485\linewidth}
      \includegraphics[height=3.05cm, width=1\linewidth]{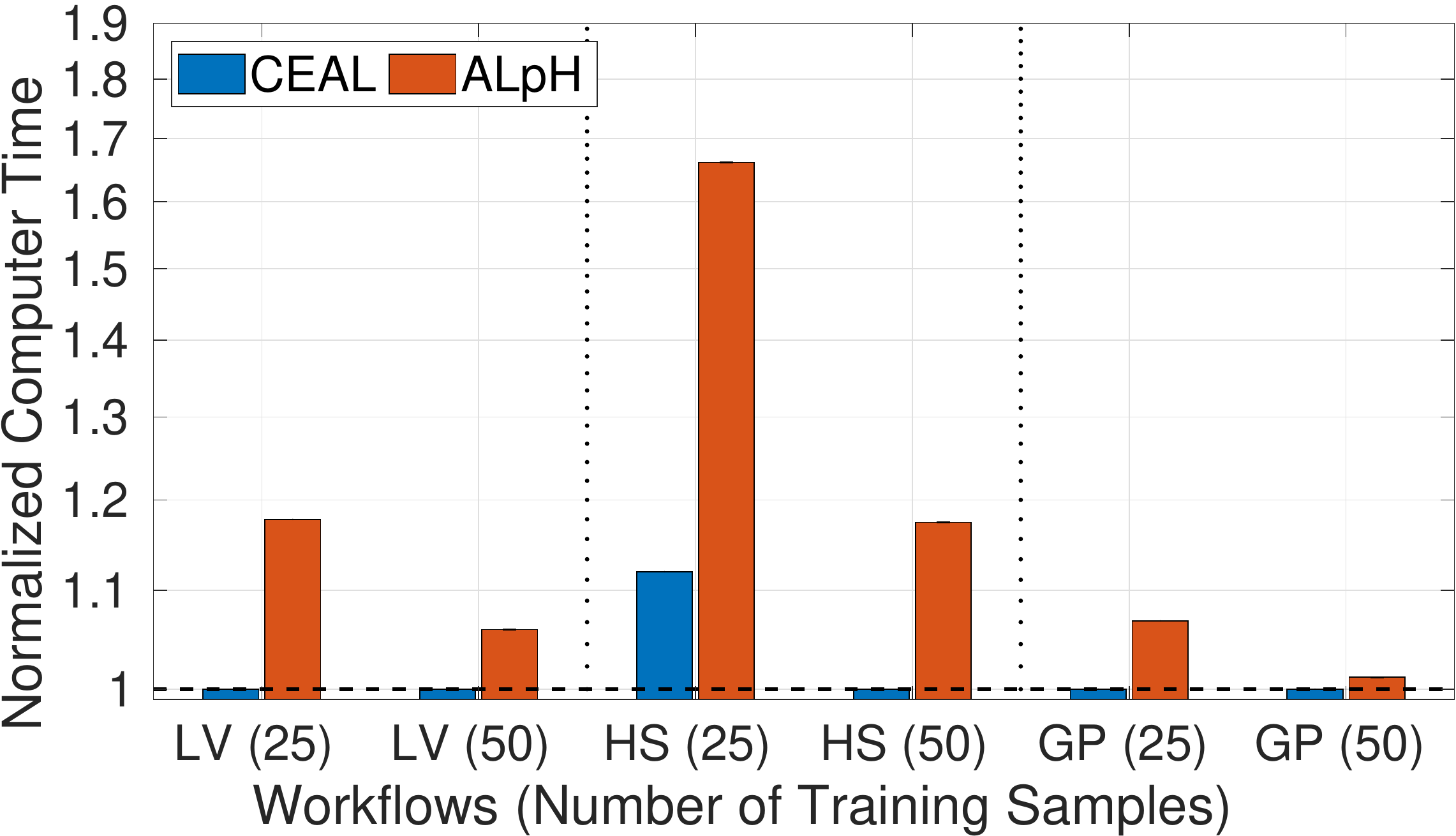}
      
            \vspace{-1ex}
            
      \caption{\it Optimizing computer time}
      \label{figure:topPerf_hist_comp}
    \end{subfigure}
    \vspace{-2.5ex}
      \caption{Best configuration auto-tuned with histories (dashed lines: best test config.)}
      \vspace{1ex}
    \label{figure:topPerf_hist}
    \begin{subfigure}[t]{.485\linewidth}
      \includegraphics[height=3.05cm, width=1\linewidth]{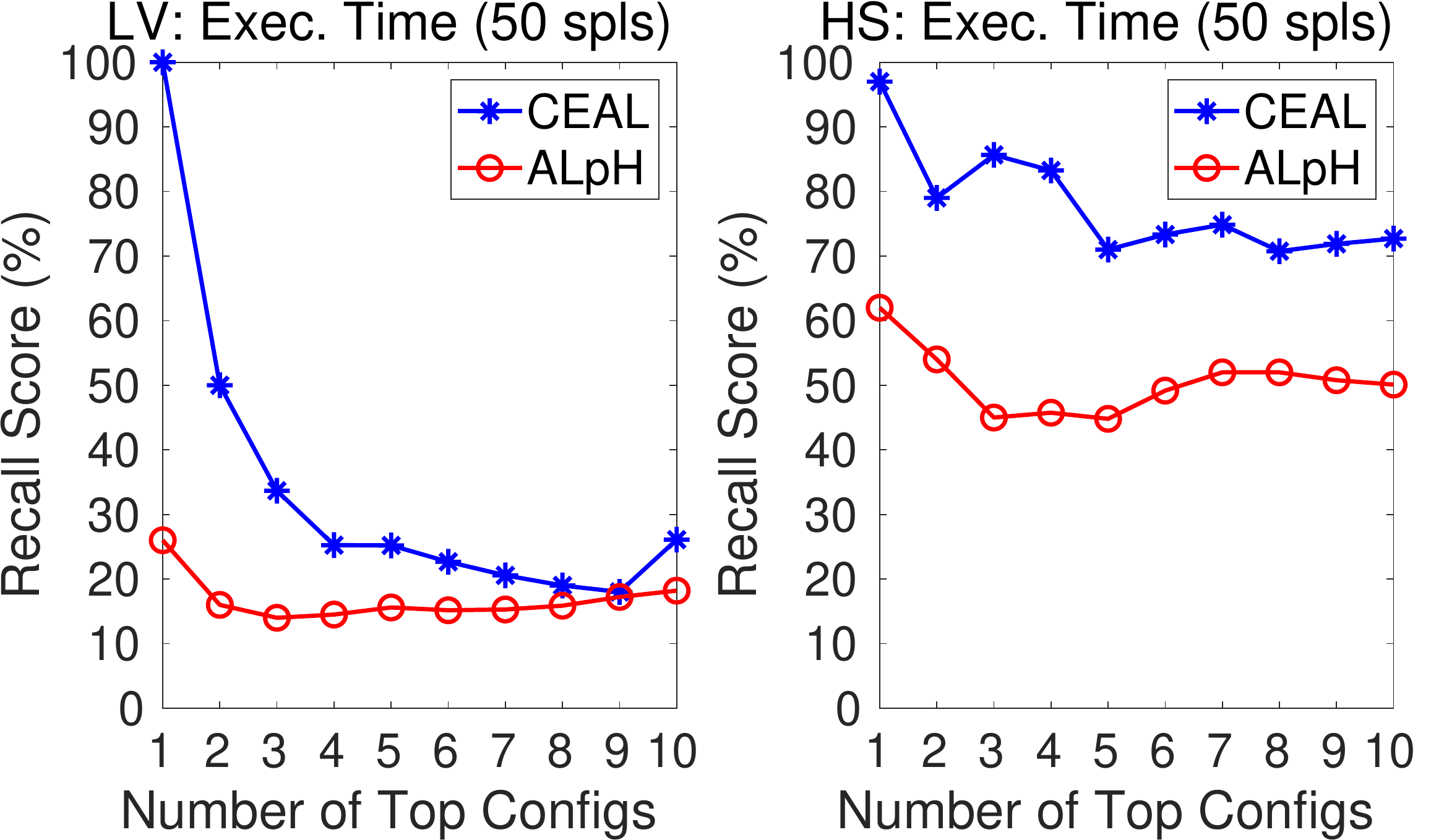}
      
            \vspace{-1ex}
            
      \caption{\it Optimizing execution time of LV and HS}
      \label{figure:rs_hist_exec}
    \end{subfigure}
    \hspace{0.005\linewidth}
    \begin{subfigure}[t]{.485\linewidth}
      \includegraphics[height=3.05cm, width=1\linewidth]{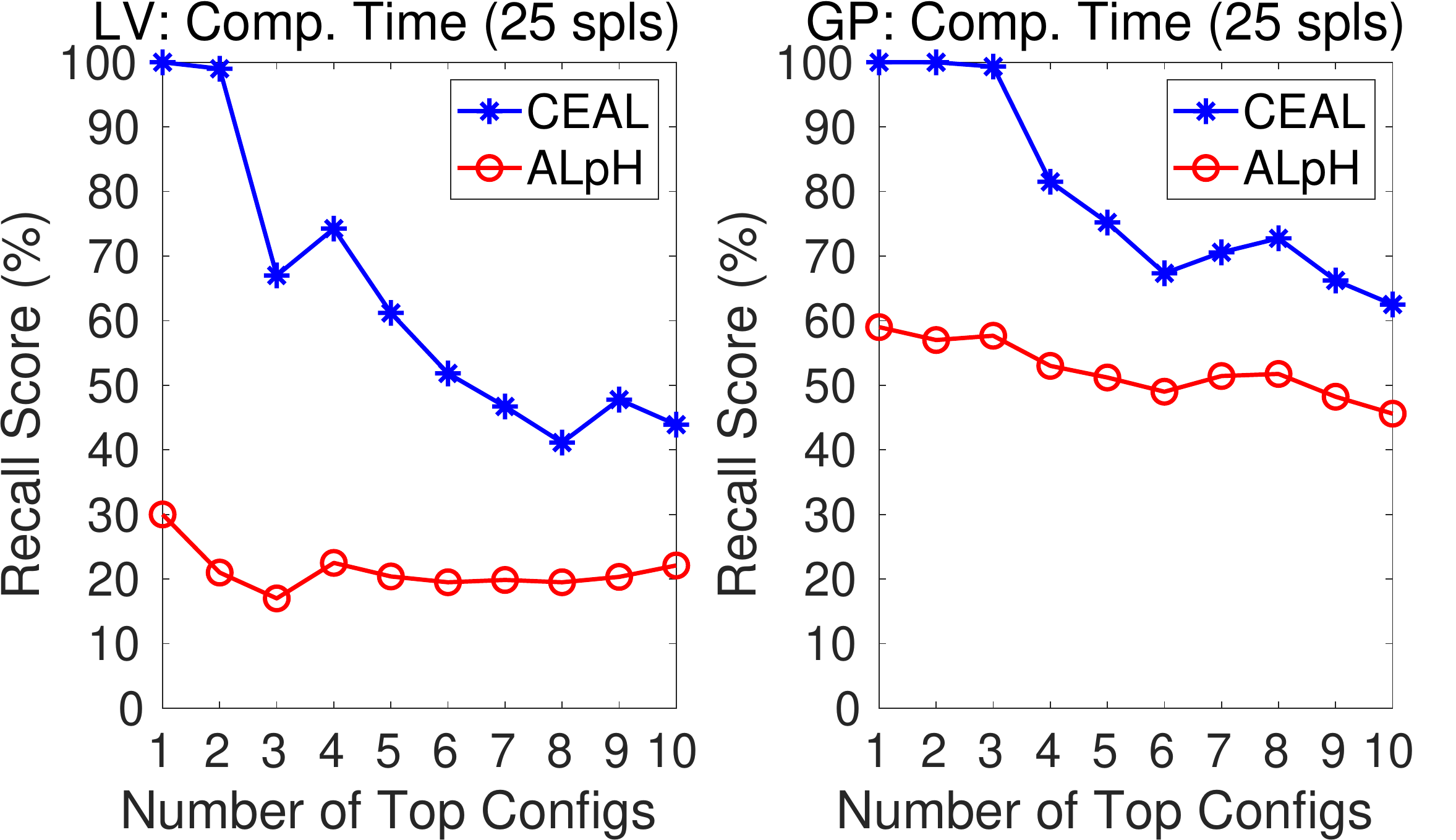}
      
      \vspace{-1ex}
      
      \caption{\it Optimizing computer time of LV and GP}
      \label{figure:rs_hist_comp}
    \end{subfigure}
    \vspace{-2.5ex}
    \caption{Robustness of autotuning with historical measurements}
    \label{figure:rs_hist}
  \end{minipage}
  \hspace{0.005\linewidth}
  \begin{minipage}[t]{.32\linewidth}
    \begin{subfigure}[t]{1\linewidth}    
      \includegraphics[height=3.05cm, width=1\linewidth]{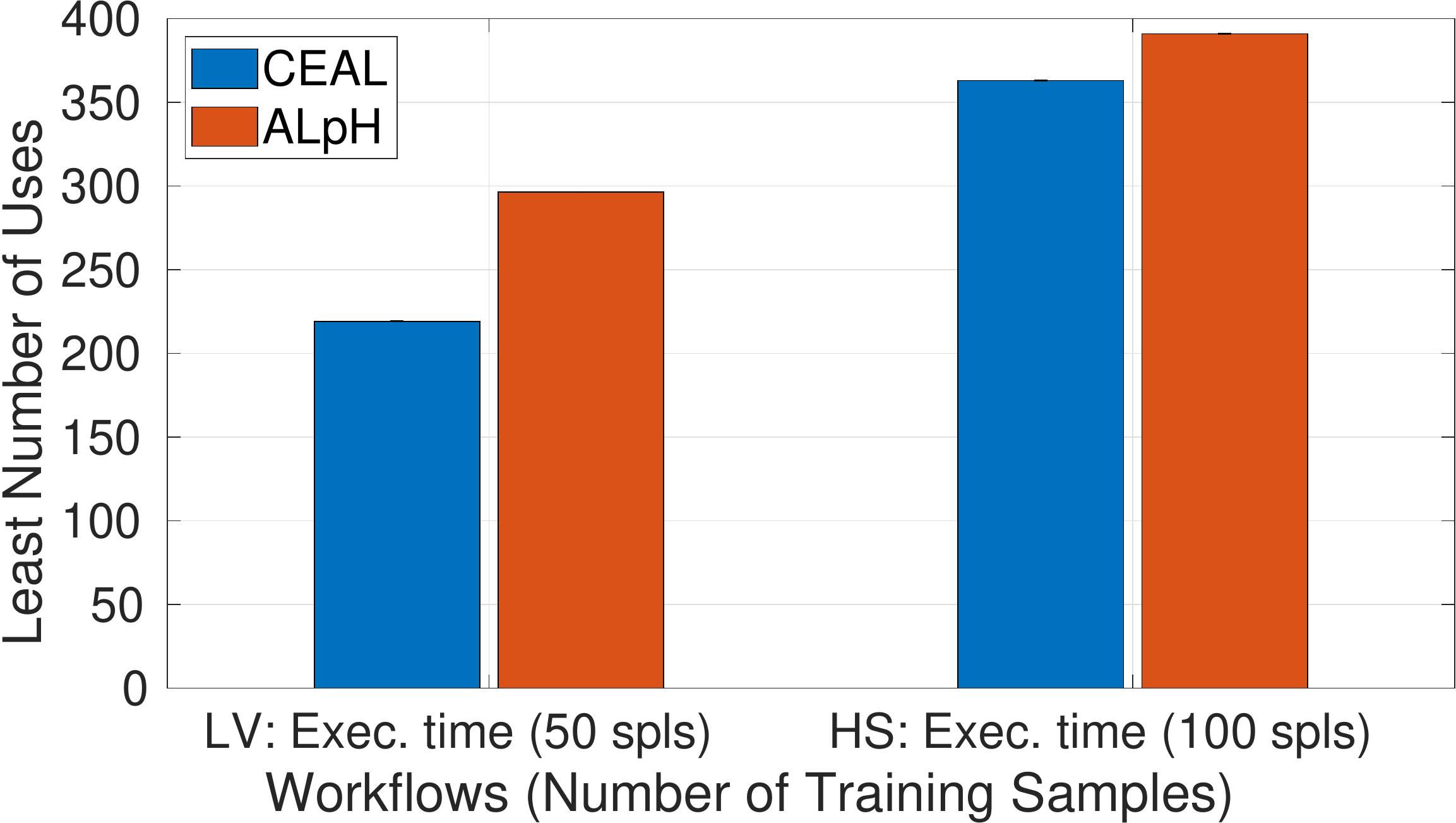}
      
            \vspace{-1ex}
            
      \caption{\it Optimizing execution time}
      \label{figure:numUse_hist_exec}
    \end{subfigure}
    \hspace{0.005\linewidth}
    \begin{subfigure}[t]{1\linewidth}
      \includegraphics[height=3.05cm, width=1\linewidth]{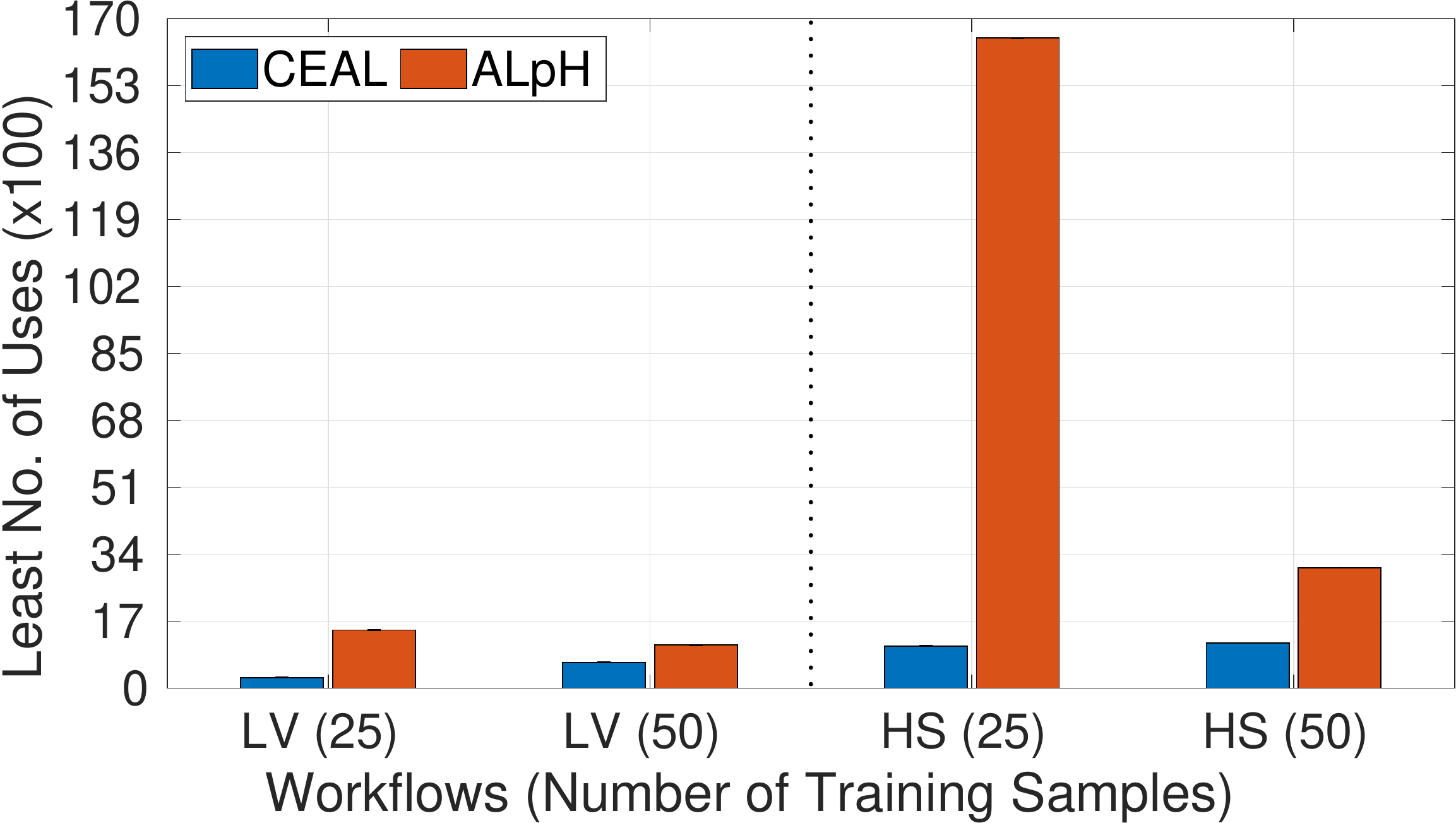}
      
            \vspace{-1ex}
            
      \caption{\it Optimizing computer time}
      \label{figure:numUse_hist_comp}
    \end{subfigure}
    \vspace{-2.5ex}
    \caption{Practicality of autotuning with historical measurements}
    \label{figure:numUse_hist}
  \end{minipage}
\end{center}
\vspace{-3ex}
\end{figure*}

Since component historical measurements are often available, 
we next examine auto-tuner performance when historical measurements 
are considered. 
We show that CEAL can make good use of component measurements 
to enhance workflow performance (\SL{Hceal}) and auto-tuner practicality (\SL{Hpractice}). 
We also demonstrate the superiority of CEAL over
ALpH's component integration method (\SL{Hperf}, \SL{Hrobust}).

\vspace{-0.05in}
\subsubsection{Effect of Previous Component Measurements}
\label{section:Hceal}
If historical performance measurements are available for component applications,
then CEAL can use those measurements to train component application models 
without charge against its training sample budget, 
and then perform more full workflow runs that would otherwise be possible.
In order to explore the benefits that result,
we compare workflow performance when optimized by CEAL with and without 
historical measurements.
In the first case, we assume no historical measurements, and thus subtract the
$m_R$ component samples used to train component models from
CEAL's training sample budget; 
in the second, we treat those measurements as historical 
and do not count them toward the cost.
We see from  \Figure{topPerf_ceal}
that historical measurements improve CEAL performance in all cases.
In addition, \Figure{topPerf_ceal_comp} shows that historical measurements
help CEAL, in the case of 25 training samples, reduce computer time
for LV by 10.0\%, HS by 38.9\%, and GP by 4.8\%.
We conclude that CEAL can make effective use of historical component measurements.

\vspace{-0.05in}
\subsubsection{Actual Performance of Auto-tuned Workflows}\label{sec:alph}
\label{section:Hperf}
Recall from \SL{combo} that ALpH differs from CEAL in using learning rather than functions
to combine component model performance predictions.
To compare these approaches to combining component models,
we measure LV, HS, and GP performance when auto-tuned by ALpH and CEAL
with different numbers of training samples.
The normalized execution and computer times in \Figure{topPerf_hist} 
show that CEAL is superior to ALpH in all cases.
For example, we see in \Figure{topPerf_hist_comp} that the
computer times of LV, HS, and GP when optimized by CEAL with 25 training samples are
15.1\%, 32.6\%, and 6.5\% less than when optimized by ALpH, respectively.
We attribute CEAL's outperforming ALpH with the same historical measurements
to CEAL's component combination using training samples more efficiently
than ALpH's model-based approach.

\vspace{-0.05in}
\subsubsection{Robustness of Auto-tuning Algorithms}
\label{section:Hrobust}
We also evaluate the robustness of ALpH and CEAL with historical component 
measurements in optimizing the execution and computer time of our in-situ workflows.
The recall scores (\SL{recallScore}) at top configurations are plotted in~\Figure{rs_hist};
we see that CEAL is always more robust than ALpH.
CEAL's best-1 and best-2 
configuration recall scores are both above 99\%.

\vspace{-0.05in}
\subsubsection{Practicality of Auto-tuning Algorithms}
\label{section:Hpractice}
We examine the practicality of ALpH and CEAL with historical component 
measurements for auto-tuning LV and HS,
and plot the number of runs (\SL{netGain}) in~\Figure{numUse_hist}.
It can be observed that when CEAL is used to optimize 
LV execution time with 50 training samples and 
LV computer time with 25 training samples,
the number of LV runs (\SL{netGain}) required to recoup the auto-tuning cost is
only 
219 and 
269, implying great practicality of CEAL.
As to the execution time cost, we consider workflow instances at training
configurations to run sequentially, even though the training data collection can
sometimes be completed in parallel.

\vspace{-0.05in}
\subsection{Hyper-parameter Sensitivity Analysis}\label{section:hp}
\vspace{-0.05in}
\label{section:paraSens}

\begin{figure}
\begin{center}
  \begin{subfigure}{0.325\linewidth}
    \includegraphics[height=2.15cm, width=1.\linewidth, trim=2 0 2 2,clip]{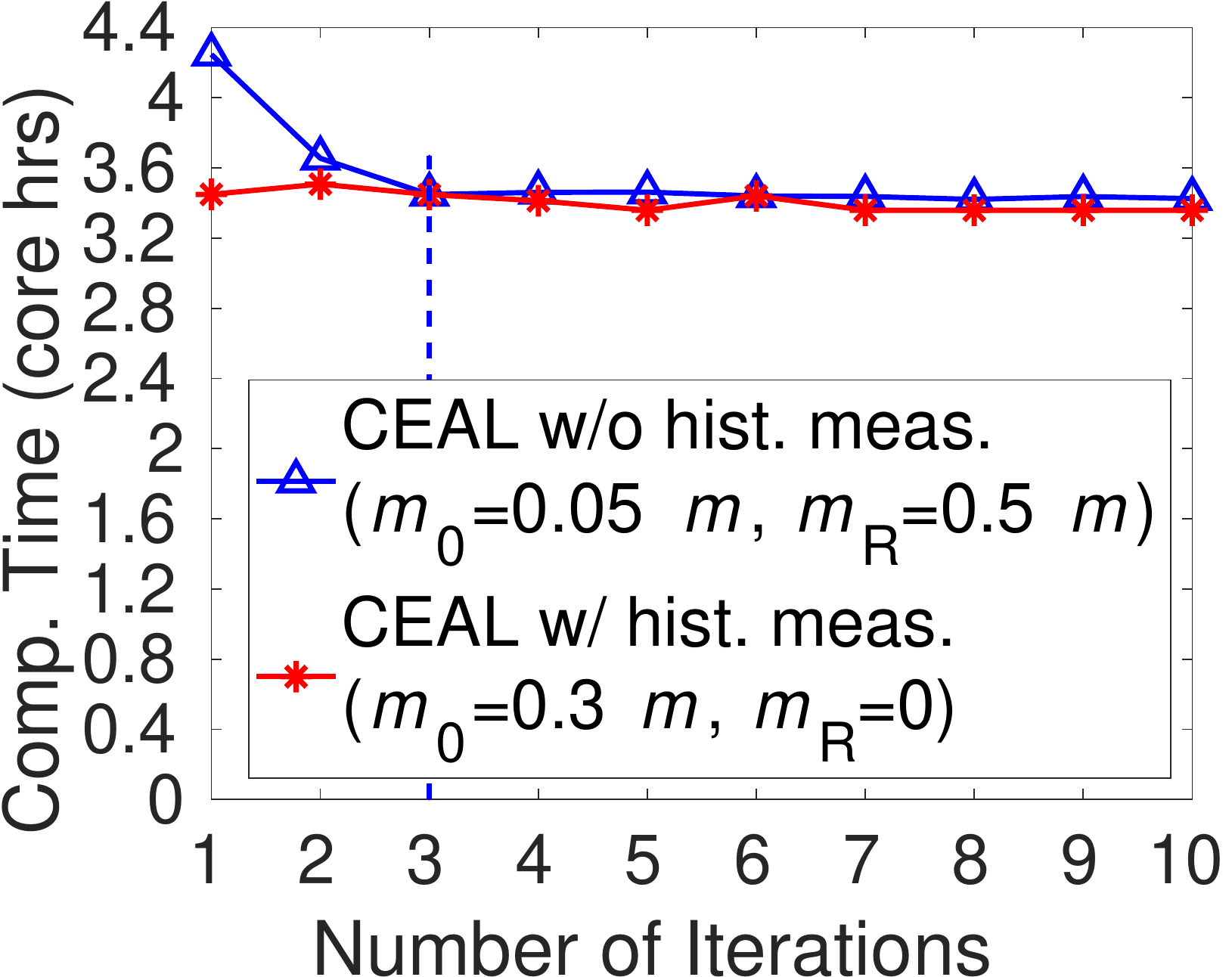}
    \caption{$I$}
    \label{figure:num_iter_comp}
  \end{subfigure}
  \begin{subfigure}{0.325\linewidth}
    \includegraphics[height=2.15cm, width=1.\linewidth, trim=2 2 2 2,clip]{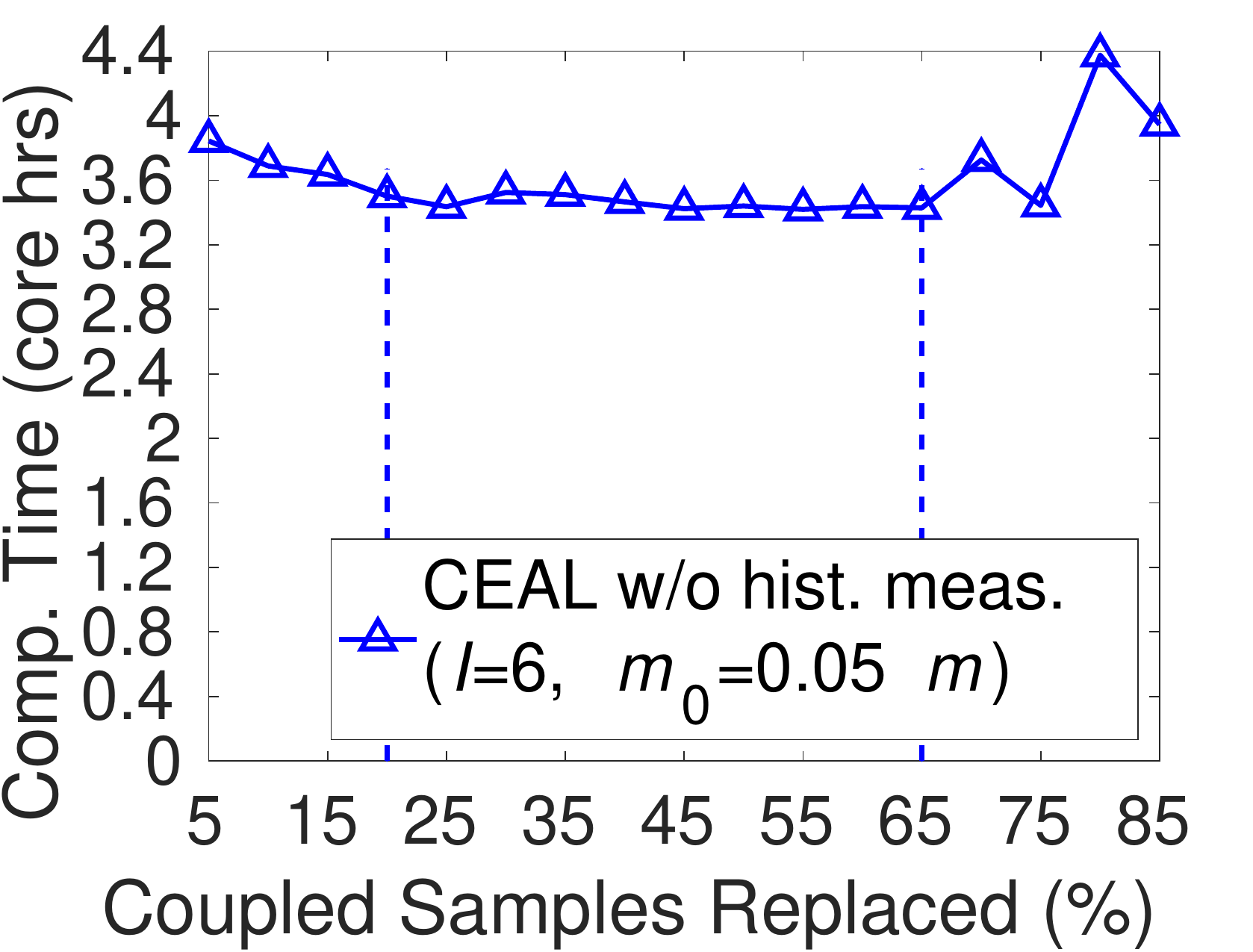}
    \caption{$m_{\R{R}}$/$m$}
    \label{figure:prec_init_comp}
  \end{subfigure}
  \begin{subfigure}{0.325\linewidth}
    \includegraphics[height=2.15cm, width=1.\linewidth, trim=2 5 2 0,clip]{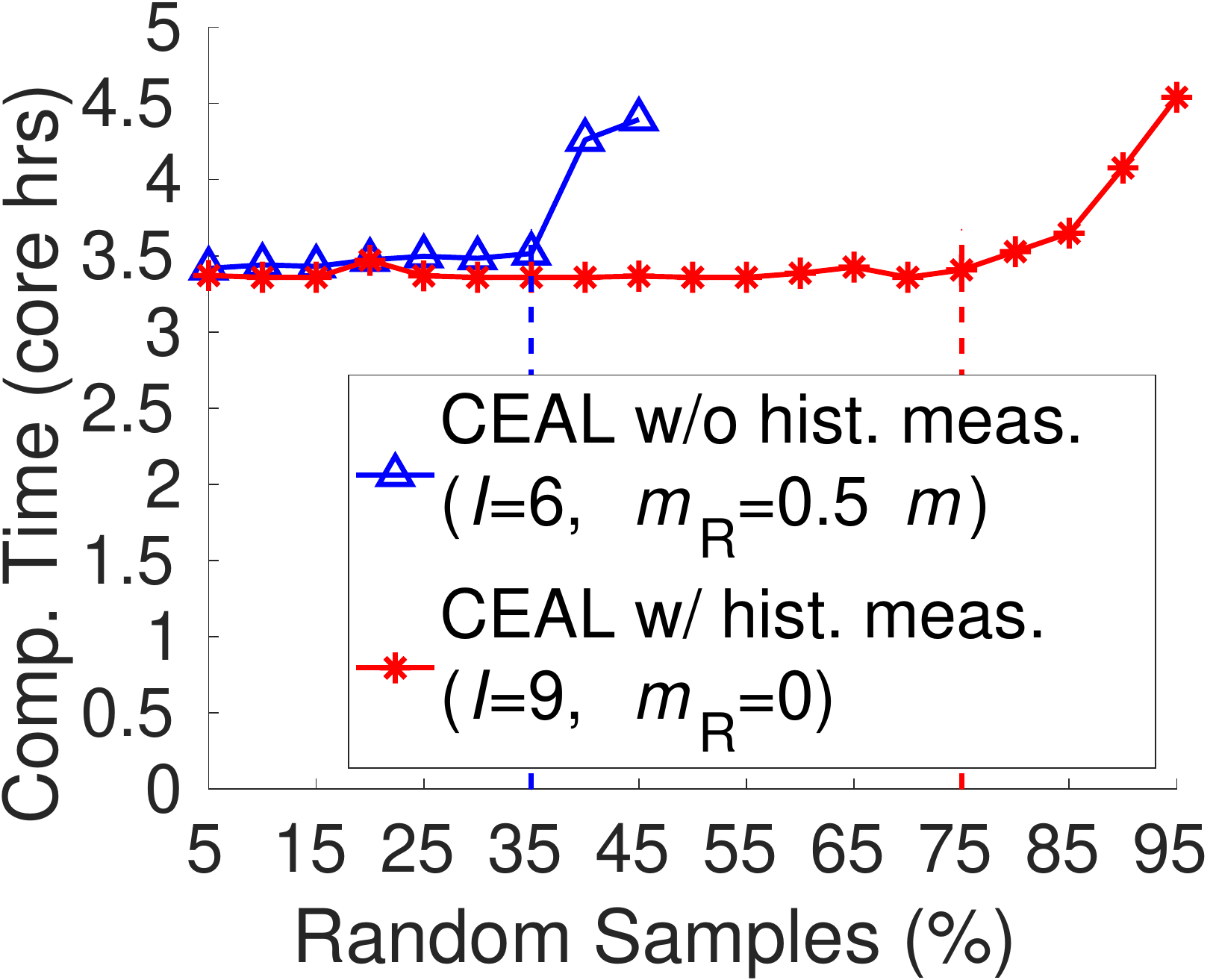}
    \caption{$m_0$/$m$}
    \label{figure:prec_rand_comp}
  \end{subfigure}
  \vspace{-3ex}
  \caption{Impact of parameter settings.}
  \label{figure:parameter}
\end{center}
\vspace{-4ex}
\end{figure}

We use LV,
when predicting the best computer time with 50 training samples,
to study CEAL's sensitivity to hyper-parameter values.
\Figure{parameter} shows the actual computer times of the best configurations predicted in various 
settings.
(1) We run CEAL with from 1 to 10 iterations ($I$). 
We see in~\Figure{num_iter_comp} that LV computer time converges to the 
best after three iterations. 
(2) We test CEAL as the number of coupled random samples 
replaced by component samples ($m_{\R{R}}$) is varied from $5\% \cdot m$ to 
$(m - m_0)$ at an interval of $5\% \cdot m$. 
\Figure{prec_init_comp} shows that LV computer time is stable over a 
large range of $m_{\R{R}}$: from 20\% to 65\% training samples.
(3) We run CEAL as we increase the number of random samples ($m_0$) 
from $5\% \cdot m$ to $(m - m_{\R{R}})$ at an interval of $5\% \cdot m$. 
\Figure{prec_rand_comp} shows that LV computer time is stable over a large 
range of $m_0$: from $5\% \cdot m$ to $35\% \cdot m$ for CEAL with 
historical measurements and from $5\% \cdot m$ to $75\% \cdot m$ for CEAL 
without historical measurements. 

We have also performed these studies for our other workflows and optimization 
metrics, and see similar results, except that in one case we found 
that an $m_0$ value of around $45\% \cdot m$ was best, indicating that 
(as discussed in \S\ref{section:actLearn}), the low-fidelity model was not 
performing well. 

\vspace{-0.05in}
\section{Related Work}
\label{section:related}
\vspace{-0.05in}

Sourouri~\etal~\cite{Sourouri:ToFinegrainDynTunHPCAppMultiCoreArch:SC17} 
integrate fine-grained auto-tuning with user-controllable hardware switches and
threads in order to use dynamic voltage and frequency scaling to improve the
energy efficiency of a memory-bound HPC application. %
However, their application-specific analytical performance model is not directly applicable to other applications.

Popov~\etal~\cite{Popov:ThrePageParaAutotunNUMA:ICS19} reduce data collection costs by
extracting and running short representative codelets, rather than whole applications, to jointly optimize page and thread mappings for applications in NUMA systems.
However,  this approach %
will not pick the optimal configuration for a whole application when that configuration is not the best of any codelet.%

Some auto-tuners target storage systems.
Cao~\etal~\cite{Cao:UnderBlackAutoTunCompAnalStorSys:ATC18}
apply and analytically compare multiple black-box optimization techiques
focusing on storage systems.
Li~\etal~\cite{Li:CAPESUnsuperviseStorTunNeuralNetReinforceLearn:SC17}
used 
reinforcement learning to develop
a model-less unsupervised storage parameter tuning system, CAPES.
However, these studies do not take data collection costs into account.

Others have applied active learning to HPC auto-tuning 
\cite{Ogilvie:MiniCostIterCompilActLearn:CGO17,
Duplyakin:ActLearnPerfAnal:CLUSTER16,
Balaprakash:ActLearnSurrMdlEmpiPerfTun:CLUSTER13}.
GEIST 
uses semi-supervised 
learning based on a parameter graph for fast parameter space 
exploration~\cite{Thiagarajan:BootParaSpacExplFastTun:ICS18};
it can auto-tune HPC applications with
configuration parameter spaces in the range of 18000.
Our work targets optimizing complex HPC applications with
configuration space size of 10$^{10}$ or more at an affordable data collection 
cost.

Marathe~\etal~\cite{Marathe:PerfMdlResConsTransLearn:SC17} 
proposed an HPC application auto-tuning algorithm that used
three fully connected neural networks 
to capture the 
relationship between configuration parameters and performance metrics 
from many cheap and a few expensive training computations.
However, the small, cheap samples often have
low similarity to the large, expensive samples, failing to provide transferable knowledge.

\vspace{-0.05in}
\section{Conclusion}
\label{section:conclusion}
\vspace{-0.05in}

The use of machine learning (ML)-based auto-tuners has been considered infeasible
for in-situ workflows due to their large configuration spaces.
Our new approach achieves high-quality ML auto-tuners for in-situ
workflows, even with a tight cost budget, by leveraging the
fact that in-situ workflows often
link multiple component applications in relatively simple
structures. 
Specifically, our CEAL algorithm 1) combines models for component
applications into a low-fidelity workflow model, and then
2) uses the low-fidelity model to guide the collection of samples for training a high-fidelity
model. Experiments with
three scientific in-situ workflows confirm the viability of the CEAL
approach, showing it can build a high-quality auto-tuner that is better at
finding best-performing
configurations than auto-tuners built with other methods.
In one example, CEAL with just 50 training samples optimizes workflow execution time
so well than only 219 subsequent runs
are required to recoup cost.

\section*{Acknowledgements}
This article reports on activities of the Codesign center for Online Data Analysis and Reduction (CODAR) \citep{foster2017computing},
which is supported by the Exascale Computing Project (17-SC-20-SC), a collaborative effort of the U.S. Department of Energy Office of Science and National Nuclear Security Administration.


%
%
                                        %
  %
  %
  %
  %
%
%
%
%
%
%
%
%
%

\bibliography{../bib/tong,../bib/autotuning,../bib/woz}

\end{document}